\documentclass[manuscript]{aastex631}

\usepackage{graphicx}	% Including figure files
\usepackage{amsmath}	% Advanced maths commands
\usepackage{amssymb}	% Extra maths symbols
\usepackage{bm}		% Bold maths symbols, including upright Greek
\usepackage{hyperref}
\usepackage{subfigure}
\usepackage{romannum}
\usepackage{color}

\usepackage{ulem}

\begin{document}

\title{Disk-wind-driven Expanding Radio-emitting Shell in Tidal Disruption Events}
%\title{Non-relativistic disk-wind-driven expanding radio-emitting shell in tidal disruption events}
%\title{Expanding radio-emitting shell originating from a disk wind in tidal disruption events}
%\title{A wind equation in tidal disruption events}

%% The \author command is the same as before except it now takes an optional
%% argument which is the 16 digit ORCID. The syntax is:
%% \author[xxxx-xxxx-xxxx-xxxx]{Author Name}
%%

\author[0000-0002-0786-7307]{Kimitake Hayasaki}
\affiliation{Department of Space Science and Astronomy, Chungbuk National University, Cheongju 361-763, Korea}
\affiliation{Harvard-Smithsonian Center for Astrophysics, 60 Garden Street, Cambridge, MA02138, USA}

\author[0000-0002-1251-7889]{Ryo Yamazaki}
\affiliation{Department of Physical Sciences, Aoyama Gakuin University, Sagamihara 252-5258, Japan}
\affiliation{Institute of Laser Engineering, Osaka University, 2-6 Yamadaoka,
Suita, Osaka 565-0871, Japan}

\begin{abstract}
We study the evolution of a non-relativistically expanding thin shell in radio-emitting tidal disruption events (TDEs) based on a one-dimensional spherically symmetric model considering the effect of both a time-dependent mass loss rate of the disk wind and the ambient mass distribution. The analytical solutions are derived in two extreme limits: one is the approximate solution near the origin in the form of the Taylor series, and the other is the asymptotic solution in which the ambient matter is dominant far away from the origin. Our numerical solutions are confirmed to agree with the respective analytical solutions. We find that no simple power-law of time solution exists in early to middle times because the mass loss rate varies over time, affecting the shell dynamics. We also discuss the application of our model to the observed radio-emitting TDE, AT2019dsg.
\end{abstract}
%
%
%%%%%%%%%%%
\section{Introduction} 
\label{sec:intro}
%%%%%%%%%%%
%

Tidal disruption events (TDEs) occur when a star passes close enough to a supermassive black hole (SMBH) and is torn apart by the SMBH tidal force \citep{hills_possible_1975}. The disrupted mass fallbacks into the SMBH \citep{rees_tidal_1988,evans_tidal_1989} and then forms an accretion disk around the SMBH due to the significant energy dissipation by debris stream-stream collisions with the help of the relativistic perihelion shift \citep{hayasaki_finite_2013,shiokawa_general_2015,bonnerot_disc_2016,hayasaki_circularization_2016}. This energy dissipation and resultant disk emissions are responsible for a flaring event with power-law decay of time over several months to years, which is observable in X-rays, ultraviolet, and optical wavebands. More than a few dozen TDE candidates have shown all or some of these signatures and radio emission has also been detected from some fraction of TDEs (e.g., see \citealt{stone_rates_2020} for a review and reference therein). 

Nonthermal radio emissions are attributed to synchrotron radiation from the relativistic electrons around shocks that are formed by interaction of relativistic or sub-relativistic outflows and surrounding material, which tells us the dynamics of the exploded ejecta (e.g., \citealt{longair_high_2011}). Sw J1644+57 is known to be the first-discovered object for jetted TDEs, which accompanies a relativistic jet viewed on-axis \citep{bloom_possible_2011,burrows_relativistic_2011,levan_extremely_2011,zauderer_birth_2011}. 
The other three jetted TDE candidates have been subsequently discovered \citep{cenko_swift_2012, brown_swift_2015, andreoni_very_2022}. 
These four jetted TDEs can also be defined as radio-loud TDEs, which are more luminous than $\sim10^{40}\,{\rm erg/s}$ in radio wavebands, while the TDEs with a luminosity less than this rough criterion are classified as radio-quiet TDEs \citep{alexander_radio_2020}. The three candidates among them are thought to have the relativistic jet but viewed off-axis \citep{lei_igr_2016, mattila_dust-enshrouded_2018, somalwar_candidate_2023}.

In recent years, several radio-quiet TDEs without any relativistic jet signatures have been observed, including ASASSN-14li \citep{alexander_discovery_2016, van_velzen_radio_2016}, XMMSL1 J0740-85 \citep{alexander_radio_2017}, CNSS J0019+00 \citep{anderson_caltech-nrao_2020}, AT2019dsg \citep{stein_tidal_2021, cendes_radio_2021}, and AT2020opy \citep{goodwin_radio_2023}. Moreover, some TDEs have shown unusual properties of radio emissions, such as a delayed flaring up of AT2018hyz \citep{cendes_mildly_2022} and a long-lived, bright flaring decay of AT2019azh \citep{goodwin_at2019azh_2022}. Either observation demonstrated that their radio emission is consistent with synchrotron radiation from shock accelerated electrons. The origin of the TDE ejecta is attributed to a disk wind \citep{dai_unified_2018}, debris stream-stream collision \citep{jiang_prompt_2016}, unbound debris stream \citep{guillochon_unbound_2016}, and so on, but it has not yet been clarified. 

For such non-relativistic radio-emitting TDEs, several theoretical models have been proposed to explain how a shocked region forms. The promising idea is that radio emission is produced by a non-relativistic wind launched from the accretion disk interacting with the circumnuclear material (e.g., \citealt{barniol_duran_radius_2013}). Alternately, an outflow is likely to be launched by debris stream-stream collision due to relativistic apsidal precession \citep{lu_self-intersection_2020}. The velocity associated with this outflow is anticipated to be slower due to the shallower gravitational potential than the disk wind case. Another possibility is that radio emission is produced by the interaction of the unbound debris stream with the ambient medium \citep{krolik_asassn-14li_2016, yalinewich_radio_2019}. The internal shock in a sub-relativistic jet is also a plausible mechanism for producing accelerated electrons. This model makes it possible to explain the observed 12-days lag between the X-ray and subsequent radio emissions in ASASSN-14li \citep{pasham_discovery_2018}. For the first scenario, the radio-emitting shell forms due to the interaction between the ambient matter and non-relativistic outflow from the TDE disk. However, the expanding motion of the shock wave front has not yet been sufficiently studied for radio-emitting TDEs.

In this paper, we newly construct a one-dimensional spherically symmetric model of the expanding shell driven by an outflow from the TDE disk, where the wind mass loss rate is assumed to follow a power-law decay of time. Our model also considers the ambient matter surrounding the shell, which has a power-law radial density profile. In Section \ref{sec:derivation}, we describe how to derive the equation of motion of the expanding shell. For confirmation purposes, we present an approximate solution around the origin and compare it with the numerical solutions. In Section~\ref{sec:longterm}, we describe the fiducial model to understand the behavior of numerical solutions and explain how the numerical solutions depend on the parameters. In Section~\ref{sec:obs}, we compare the numerical solution with the observed data of AT2019dsg \citep{cendes_radio_2021}, which is the best example for our model. Section~\ref{sec:disc} discusses the results of our model, and Section~\ref{sec:cons} is devoted to our conclusions.

%
%
%%%%%%%%%%%%%%%%%%%%%%%%%%
\section{Derivation of the expanding thin shell equation}
\label{sec:derivation}
%%%%%%%%%%%%%%%%%%%%%%%%%%
%
% 
We consider the evolution of the expanding thin shell throughout this paper, following the prescription of \citet{chevalier_radio_1982}, where the effect of both the mass addition of the ejecta to the shell and the dragging due to the surrounding matter on the motion of the shell are considered. We also neglect the gravitational force in the equation of motion for the expanding shell because the disk wind blows freely from the SMBH gravity.

In this section, we derive the equation of motion of the radius of the expanding thin shell driven by the TDE disk winds in non-relativistic radio-emitting TDEs based on the assumption that the shell width is much thinner than the shell radius (i.e., the thin shell approximation), where we set the thin shell forms at $t=t_0$. In our scenario, a star is tidally disrupted by a SMBH, and subsequently, an accretion disk forms at $t<t_0$.

%
%%%%%%%%%%%%%%%%%%%%%%%%%%%%%%%%
\subsection{Mass conservation of the shell}
\label{sec:mom}
%\subsection{Ejecta and ambient masses to form the shell}
%%%%%%%%%%%%%%%%%%%%%%%%%%%%%%%%
%
Considering the wind mass emitted from the disk at any retarded time $t^\prime$, that is between $t_0$ and $t$, where the wind's mass element leaving from $r=r_0$ at $t=t^\prime$ reaches radius $r$ at $t$. Assuming the disk wind starts blowing from $r=r_0$ in the free expansion with velocity $v_0$, we obtain
\begin{eqnarray}
&&
r=r_0+v_0(t-t^\prime).
\label{eq:streamline}
\end{eqnarray} 
The farthest radius that the wind's mass element can reach is then estimated to be 
\begin{equation}
r_{\rm max}=r_0+v_0(t-t_0).
\label{eq:rmax}
\end{equation}
Since the mass emitted at $t=t^\prime+\Delta{t}^\prime$ reaches the radius $r-\Delta{r}$, we obtain the following relation using equation~(\ref{eq:streamline}) as
\begin{eqnarray}
\Delta{r}=v_0\Delta{t^\prime}.
\label{eq:deltat}
\end{eqnarray}
We also rewrite equation~(\ref{eq:streamline}) as
\begin{equation}
t^\prime=t-\frac{r-r_0}{v_0}.
\label{eq:tret}
\end{equation}
The thin shell is thought to form at $t=t_0$ due to the interaction between the ambient matter and the disk wind, and subsequently, the disk wind continues to impact the shell dynamically. Therefore, we postulate the mass loss rate of the disk wind follows the power-law decay with time after the shell formation, i.e., for $t\ge{t}_0$ as
\begin{equation}
    \dot{M}_{\rm w}(t)=\dot{M}_0\left(\frac{t}{t_0}\right)^{-n},
    \label{eq:mdotw}
\end{equation}
where $n$ is the power-law index and, e.g., $n=5/3$ corresponds to the power-law index of the mass fallback rate \citep{evans_tidal_1989} and note that, in this case, the mass loss rate is proportional to the mass fallback rate \citep{strubbe_optical_2009}. Since the outflowing mass is conserved on the spacetime diagram as
\begin{equation}
    \dot{M}_{\rm w}(t^\prime)\Delta{t^\prime}=4\pi{r}^2\Delta{r}\rho_{\rm ej}(r,t),
    \nonumber 
\end{equation}
we obtain the unshocked ejecta's mass density as
\begin{equation}
\rho_{\rm ej}(r,t)
=\rho_{\rm ej,0}
\left(\frac{r}{r_0}\right)^{-2}
\left(
\frac{t}{t_0}-\frac{r-r_0}{v_0t_0}
\right)^{-n},
\label{eq:rhoej}
\end{equation}
where equations~(\ref{eq:deltat})-(\ref{eq:mdotw}) are adopted for the derivation and $\rho_{\rm ej,0}$ is defined as the initial ejecta density by  
\begin{equation}
\rho_{\rm ej,0}\equiv\frac{\dot{M}_0}{4\pi{r_0^2}v_0}.
\label{eq:rhoej0}
\end{equation}
The ejecta mass comprising the infinitesimally thin shell at radius $R(t)$ is then obtained by using equations (\ref{eq:rmax}) and (\ref{eq:rhoej}) as
\begin{eqnarray}
M_{\rm ej}(t)
&&
=
\int_{R(t)}^{r_{\rm max}}\,4\pi{r^2}\rho_{\rm ej}(r,t)\,dr
\nonumber \\
&&
=\frac{\dot{M}_0t_0}{n-1}
\Biggr[
1-
\left(
\frac{r_0+v_0{t}-R(t)}{v_0t_0}
\right)^{1-n}
\Biggr].
\label{eq:mej}
\end{eqnarray}
The time derivative of equation~(\ref{eq:mej}) is given by
\begin{eqnarray}
\dot{M}_{\rm ej}(t)
&&
\equiv
\frac{d{M}_{\rm ej}(t)}{dt}
\nonumber \\
&&
=\dot{M}_0
\left(
1-\frac{\dot{R}(t)}{v_0}
\right)
\left[
\frac{r_0+v_0{t}-R(t)}{v_0t_0}
\right]^{-n}.
\label{eq:mdotej}
\end{eqnarray}
This is positive because $R(t)<v_0t$ and $\dot{R}(t)<v_0$, meaning that the ejecta mass increases with time.

We assume that the ambient matter density is given by
\begin{equation}
\rho_{\rm am}=\rho_{\rm am,0}\left(\frac{r}{r_0}\right)^{-s},
\label{eq:rhoam}
\end{equation}
where $s$ is the power-law index. Integrating it from $r_0$ to $R(t)$, we obtain the shell mass, to which is contributed from the ambient matter, as
\begin{equation}
M_{\rm am}(t)=\frac{4\pi\rho_{\rm am,0}r_0^3}{3-s}\left(\frac{R(t)}{r_0}\right)^{3-s}\left[1-\left(\frac{r_0}{R(t)}\right)^{3-s}\right].
\label{eq:amm}
\end{equation}
The time derivative of $M_{\rm am}(t)$ is given by
\begin{equation}
\dot{M}_{\rm am}(t)
\equiv
\frac{dM_{\rm am}(t)}{dt}
=
4\pi\rho_{\rm am,0}r_0^3\left(\frac{R(t)}{r_0}\right)^{2-s}\frac{\dot{R}(t)}{r_0}.
\label{eq:mdotam}
\end{equation}
Now we introduce $\delta$ as a parameter to decide the amount of the initial shell mass $\Delta{m}$:   
\begin{equation}
\delta=\frac{\Delta{m}}{4\pi\rho_{\rm ej,0}r_0^3}.
%=\delta\epsilon\dot{M}_0t_0=\delta{F}_{\rm ej,0}\left(\frac{V_0}{t_0}\right)^{-1},
\label{eq:inimass} 
\end{equation}
The total mass of the thin shell at radius $R(t)$ is then given by
\begin{eqnarray}
M(t)
&&
=
M_{\rm ej}(t)+M_{\rm am}(t)+\Delta{m}
\nonumber \\
&&
=
\frac{\dot{M}_0t_0}{n-1}
\Biggr[
1-
\left(
\frac{r_0+v_0{t}-R(t)}{v_0t_0}
\right)^{1-n}
+
(n-1)
\delta\epsilon
\Biggr]
+
\frac{4\pi\rho_{\rm am,0}r_0^3}{3-s}\left(\frac{R(t)}{r_0}\right)^{3-s}\left[1-\left(\frac{r_0}{R(t)}\right)^{3-s}\right],
\nonumber \\
\label{eq:totm}
\end{eqnarray}
where the parameter $\epsilon$ is introduced as a dimensionless quantity to determine the initial wind velocity
\begin{equation}
V_0\equiv\frac{r_0}{t_0},
\label{eq:v0}
\end{equation}
defined such that 
\begin{eqnarray}
\epsilon
&&
\equiv
\frac{V_0}{v_0}.
\label{eq:epsilon0}
\end{eqnarray}

%
%%%%%%%%%%%%%%%%%%%%%%%%%%%%%%%%%%%%%%%%%%%%
\subsection{Equation of motion of the shell}
\label{sec:eom}
%%%%%%%%%%%%%%%%%%%%%%%%%%%%%%%%%%%%%%%%%%%%
%
Let us consider the momentum conservation law of the system composing of the ejecta, shell, and ambient matter between $t$ and $t+\Delta{t}$. The momentum conservation law is given by 
\begin{eqnarray}
\Delta{M}_{\rm ej}(t)v_0 + M(t)\dot{R}(t)
&&
=M(t+\Delta{t})\dot{R}(t+\Delta{t}), 
\label{eq:momcl}
\end{eqnarray}
where $\Delta{M}_{\rm ej}(t)$ is the mass of the ejecta added by colliding with the shell during $\Delta{t}$. 
Ignoring the terms more than the second-order of $\Delta{t}$ in the Taylor series of the right-hand side finally yields the equation of motion at the limit of $\Delta{t}\rightarrow0$ as
\begin{eqnarray}
 M(t)\ddot{R}(t)=F(t),
 \label{eq:eom}
 \end{eqnarray}
where 
\begin{eqnarray}
F(t)
&&
\equiv
F_{\rm ej}(t) - F_{\rm am}(t)
\nonumber \\
&&
=
\frac{F_{\rm ej,0}}{V_0^2}
\left[
\frac{r_0+v_0{t}-R}{v_0t_0}
\right]^{-n}
    \left(
v_0-\dot{R}(t)
\right)^2
-
\frac{F_{\rm am,0}}{V_0^2}
\left(\frac{R}{r_0}\right)^{2-s}\dot{R}(t)^2
\nonumber 
\\
\label{eq:force}
\end{eqnarray}
with $F_{\rm ej}(t)\equiv4\pi{R^2}P_{\rm ej}$ and $F_{\rm am}(t)\equiv4\pi{R^2}P_{\rm am}$.
In this context, $P_{\rm ej}=\rho_{\rm ej}(v_0-\dot{R}(t))^2$ and $P_{\rm am}=\rho_{\rm am}\dot{R}(t)^2$ signify the ram pressures of the ejecta and the ambient matter, respectively. Furthermore, $F_{\rm ej,0}$ and $F_{\rm am,0}$ are denoted by
%Here, $P_{\rm ej}=\rho_{\rm ej}(v_0-\dot{R}(t))^2$ and $P_{\rm am}=\rho_{\rm am}\dot{R}(t)^2$ are the ejecta's and ambient matter's ram pressures, respectively, and $F_{\rm ej,0}$ and $F_{\rm am,0}$ are defined by
\begin{eqnarray}
F_{\rm ej,0}
&&
\equiv
4\pi\rho_{\rm ej,0}{r_0}^2V_0^2=\epsilon\dot{M}_0V_0,
\label{eq:fej0}
\\
F_{\rm am,0}
&&
\equiv
4\pi\rho_{\rm am,0}\,r_0^2V_0^2=\eta\epsilon\dot{M}_0V_0
\label{eq:fam0}
\end{eqnarray}
with the subsequent introduction of a dimensionless parameter:
\begin{eqnarray}
\eta
\equiv
\frac{F_{\rm am,0}}{F_{\rm ej,0}}=\frac{\rho_{\rm am,0}}{\rho_{\rm ej,0}},
\label{eq:eta0}
\end{eqnarray}
where equations~(\ref{eq:rhoej0}) and (\ref{eq:v0}) are adopted. Note that equation~(\ref{eq:eom}) with equation~(\ref{eq:force}) is consistent with the equation of motion of the expanding thin shell
\citep{chevalier_radio_1982}.
%
%
%%%%%%%%%%%%%%%%%%%%
% Non-dimensional equation of motion
%%%%%%%%%%%%%%%%%%%%
%
Substituting equation (\ref{eq:totm}) into equation~(\ref{eq:eom}) with equations~(\ref{eq:force}) to (\ref{eq:eta0}), we finally obtain the dimensionless equation of motion of the shell as
\begin{eqnarray}
\ddot{y}
&&
=
\Biggr[
(n-1)
\left[
x+\epsilon(1-y)
\right]^{-n}
\left[
2\dot{y}
-
\epsilon\dot{y}^2
-
\frac{1}{\epsilon}
\right]
+
\eta
\epsilon(n-1)
y^{2-s}
\dot{y}^2
\Biggr]
\nonumber \\
&&
\biggr{/}
\Biggr[
\left[
x+\epsilon(1-y)
\right]^{1-n}
-
\frac{
\eta\epsilon(n-1)
}{3-s}
(y^{3-s}-1)
-
\delta\epsilon
(n-1)
-1
\Biggr],
\label{eq:eom2}
\end{eqnarray}
where we adopt the following dimensionless parameters:
\begin{align}
  \begin{split}
  x&\equiv\frac{t}{t_{\rm 0}}, \quad \\
y&\equiv\frac{R(t)}{r_0}, \quad \\
\dot{y}&\equiv\frac{dy}{dx}=\frac{\dot{R}(t)}{V_0}, \quad \\
\ddot{y}&\equiv\frac{d^2y}{dx^2}=\ddot{R}(t)\frac{t_0}{V_0}. \quad 
%    ds^2 = - A(r) dt^2 + B(r) dr^2 + r^2 (d\theta^2 + \sin^2\theta d\phi^2), \\
%    0 \le r \le \infty, \quad
 %   0 \le \theta < \pi, \quad
%    0 \le \phi < 2\pi
  \end{split}
 \label{eq:dlq} 
\end{align}
We solve equation~(\ref{eq:eom2}) using the Runge-Kutta method with the initial conditions: $x_0\equiv{x}(t_0)=1$, $y_0\equiv{y}(R(t_0))=y(r_0)=1$, and $\dot{y}(\dot{R}(t_0))=\dot{y}(v_0)=1/\epsilon$.

%
%
%%%%%%%%%%%%%%%%%%%%%%%%
\subsection{Comparison to analytical solutions near the wind source}
%$r_0$}
\label{sec:analsol} 
%%%%%%%%%%%%%%%%%%%%%%%%
%
%
%%%%%%%%%
% Figure
%%%%%%%%%
%
\begin{figure}[ht!]
\centering
\subfigure[]{\includegraphics[scale = 0.44]{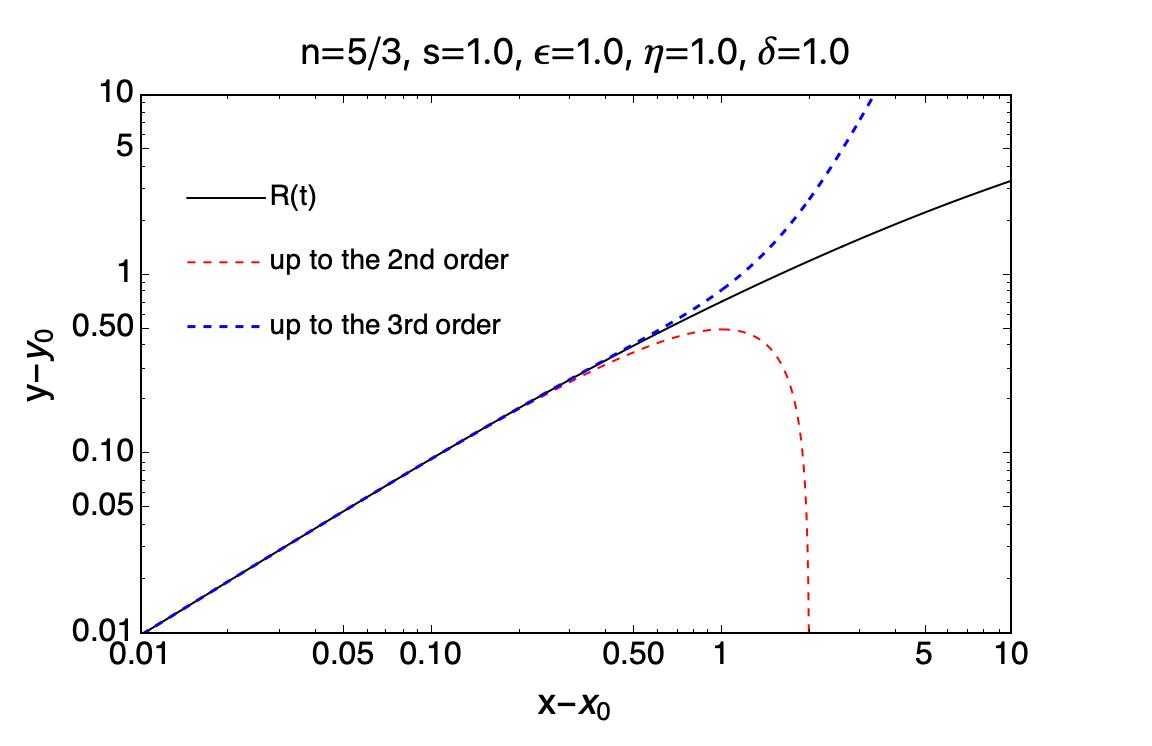}}
\subfigure[]{\includegraphics[scale = 0.44]{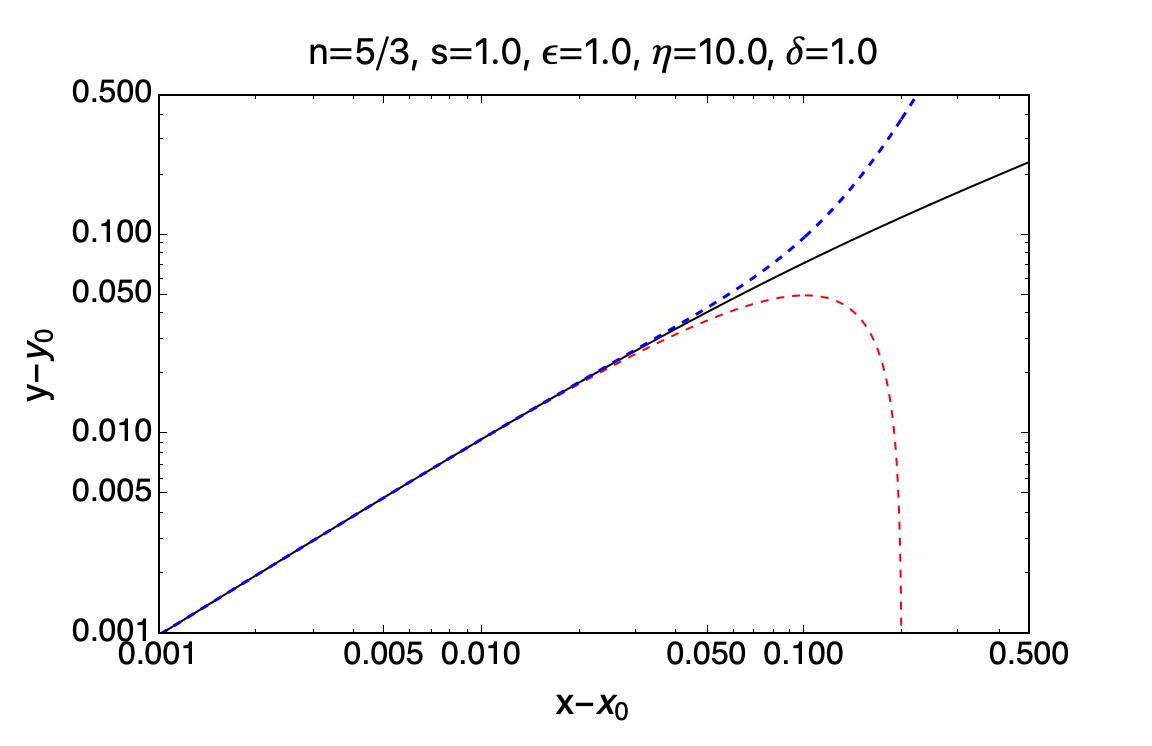}}\\
\subfigure[]{\includegraphics[scale = 0.44]{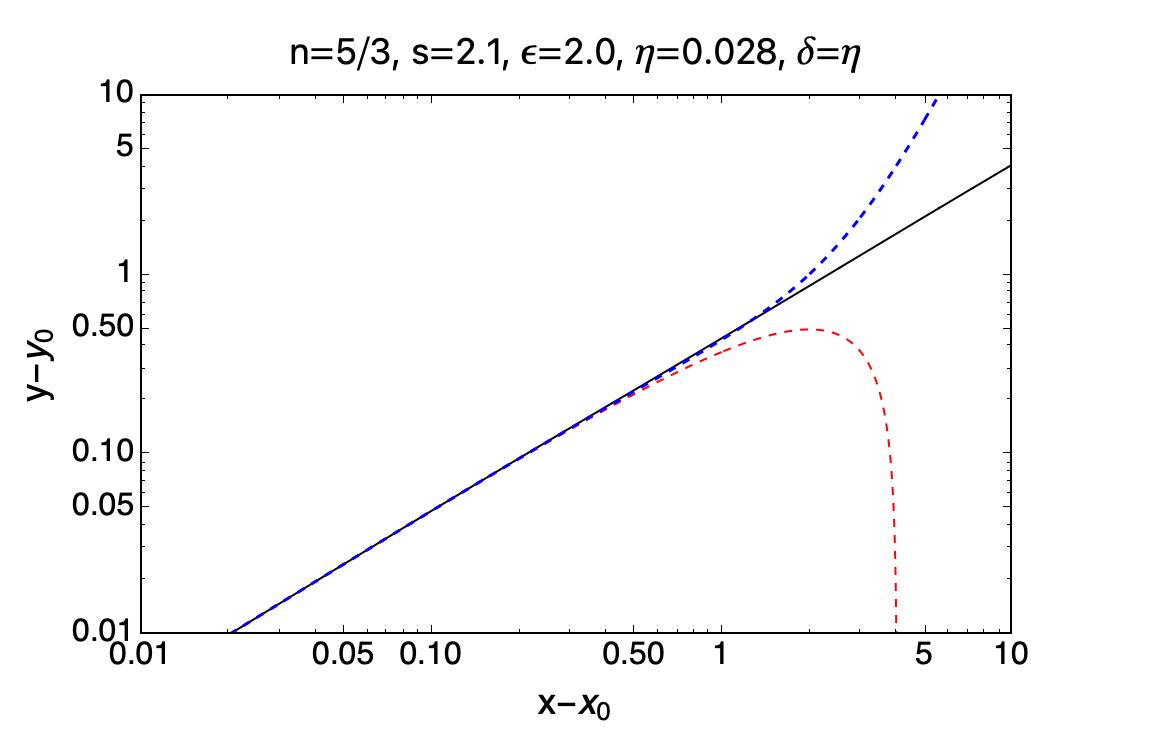}}
\subfigure[]{\includegraphics[scale = 0.44]{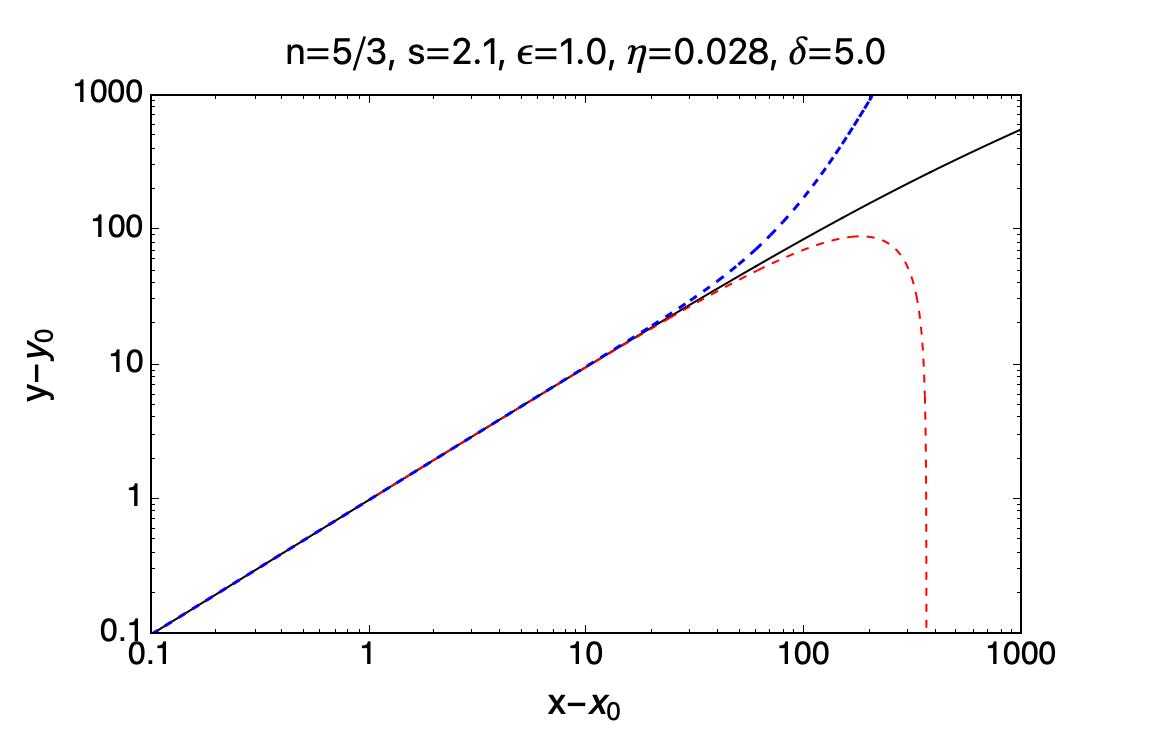}}
\caption{
Comparison of the dimensionless expanding shell radius $y(x)$ between analytical and numerical solutions for different parameters. 
{
The vertical and horizontal axes denote $y-y_0$ and $x-x_0$, respectively, where $y=R(t)/r_0$, $y_0=1$, $x=t/t_0$, and $x_0=1$. In each panel,} the solid black line represents the numerical solution of the shell radius, while the red and blue dashed lines show the approximate solutions given by equation~(\ref{eq:approx}) up to the second and third orders, respectively. 
}  
\label{fig:cana}
\end{figure}

In this section, we confirm whether we numerically solve equation~(\ref{eq:eom2}) correctly compared to an approximate solution near the wind source. We derive the approximate solution in the form of the Taylor series (see the detail in Appendix~\ref{sec:app}) as
\begin{equation}
y(x)
\approx
1 + \frac{1}{\epsilon}(x-x_0) - \frac{1}{2}\frac{\eta}{\epsilon^2\delta}(x-x_0)^2  + \frac{1}{6}\frac{\eta}{\epsilon^3\delta}
\left(
3\frac{\eta}{\delta}
+
s-2
\right)
(x-x_0)^3.
\label{eq:approx}
\end{equation}
It is found that equation~(\ref{eq:approx}) shows no dependence on $n$ up to the third order and depends only on $\epsilon$ at the leading term. Figure~\ref{fig:cana} compares the numerical solutions for equation~(\ref{eq:eom2}) with equation~(\ref{eq:approx}) with different parameters. In panel (a), we adopt $(n,\,s,\,\epsilon,\,\eta,\,\delta)=(5/3,\,1.0,\,1.0,\,1.0,\,1,0)$ for both numerical and approximate solutions as the fiducial case. The same parameters are adopted for panel (b) as those of panel (a) but for $\eta=10$. In panels (c) and (d), we apply the following parameter sets {$(n,\,s,\,\epsilon,\,\eta,\,\delta)=(5/3,\,2.1,\,2.0,\,0.028,\,0.028)$ and $(5/3,\,2.1,\,1.0,\,0.028,\,5.0)$}, respectively. These sets are introduced for two purposes: one is to check whether a qualitative difference appears when we compare both numerical and approximate solutions by adopting far different parameters from the fiducial case, and the other reason is that we adopted one of the parameter sets for comparing with the observational data of AT2019dsg (see Section~\ref{sec:compobs} for the detail). From these four panels, we demonstrate the approximate solution is in good agreement with the numerical solutions around $x=x_0$ as predicted, ensuring the validity of numerical solutions clearly.

%
%%%%%%%%%%%%%%%%%%%%%%%%%%%%
\section{Longterm evolution of the expanding thin shell}
\label{sec:longterm}
%%%%%%%%%%%%%%%%%%%%%%%%%%%%
%
In this section, we describe the longterm evolution of physical quantities to characterize the expanding thin shell and how the five parameters $(n,\,s,\,\epsilon,\,\eta,\,\delta)$ change them. 
%the ejecta and ambient masses, their changing rates, force acting on each matter, and local power law index of the shell radius, 
%
%%%%%%%%%
% Figure
%%%%%%%%%
%
\begin{figure}[ht!]
\centering
\subfigure[]{\includegraphics[scale = 0.44]{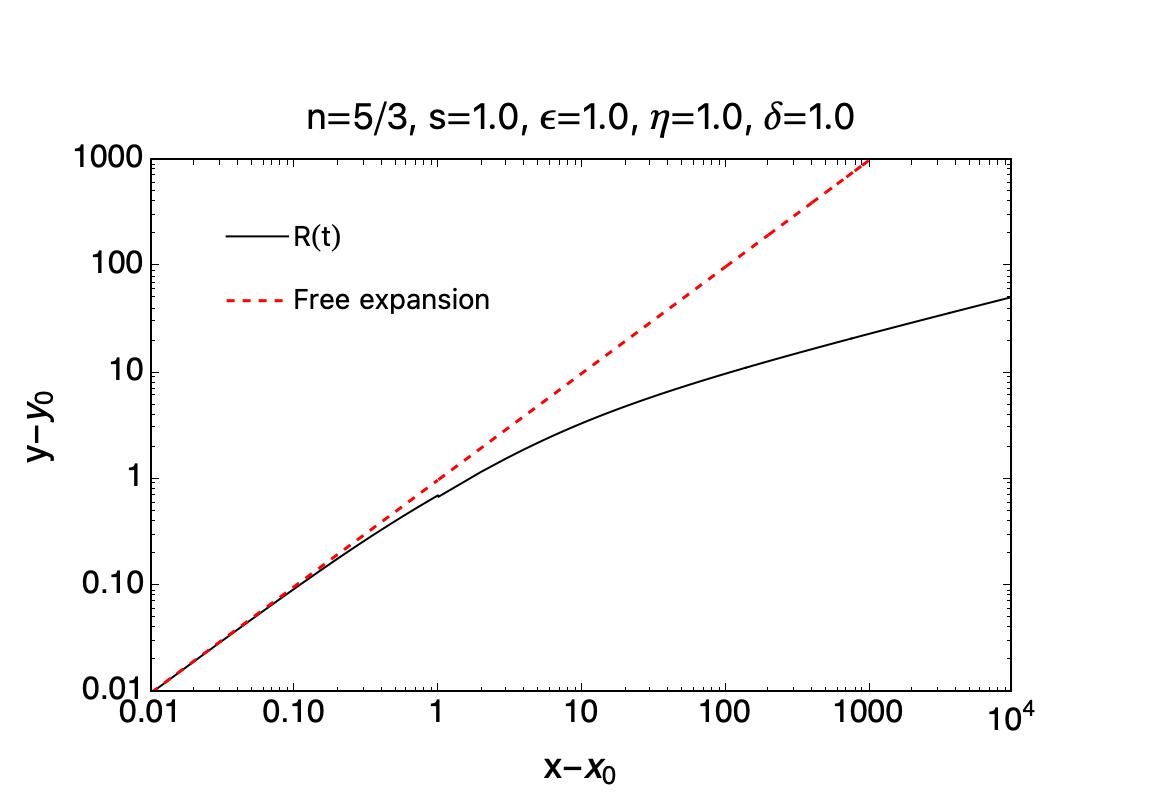}}
\subfigure[]{\includegraphics[scale = 0.44]{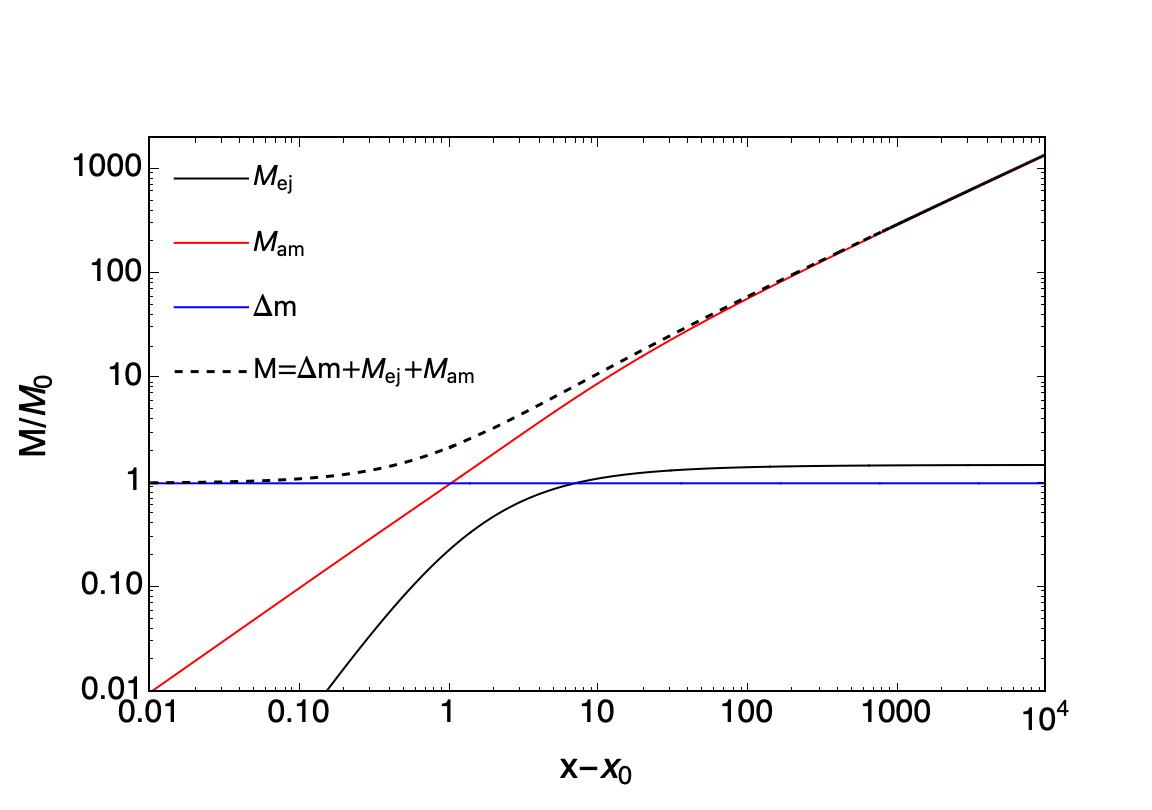}}
\subfigure[]{\includegraphics[scale = 0.44]{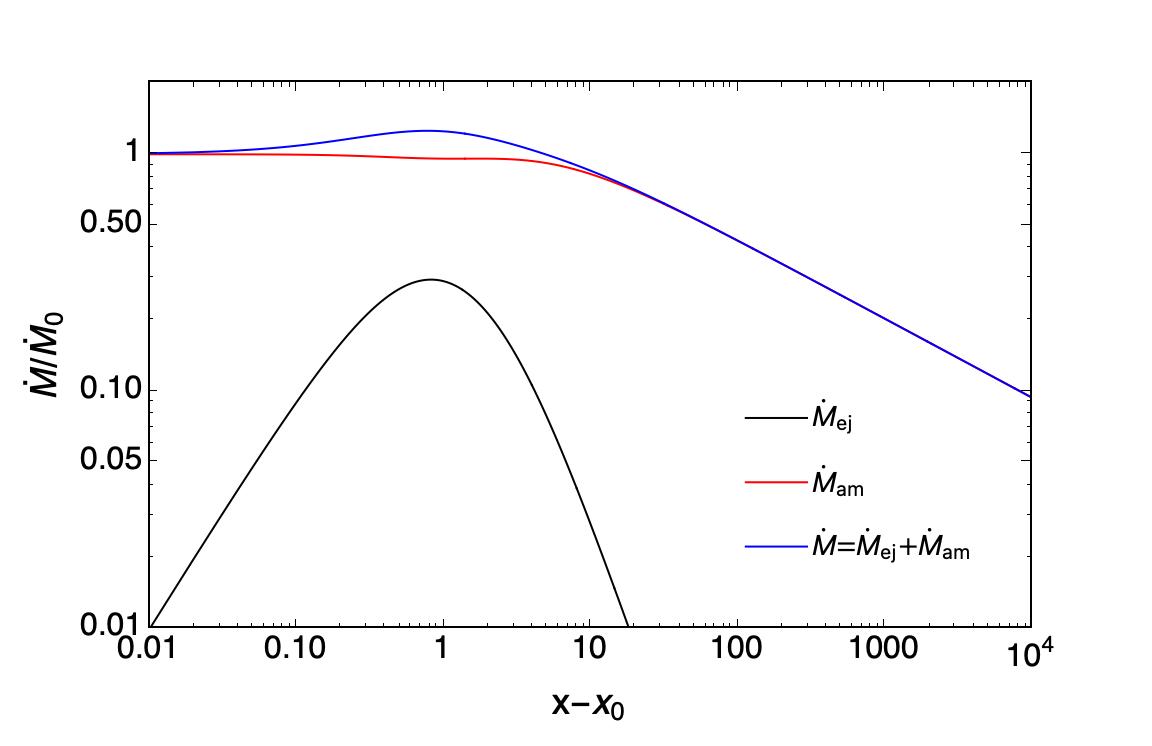}}
\subfigure[]{\includegraphics[scale = 0.44]{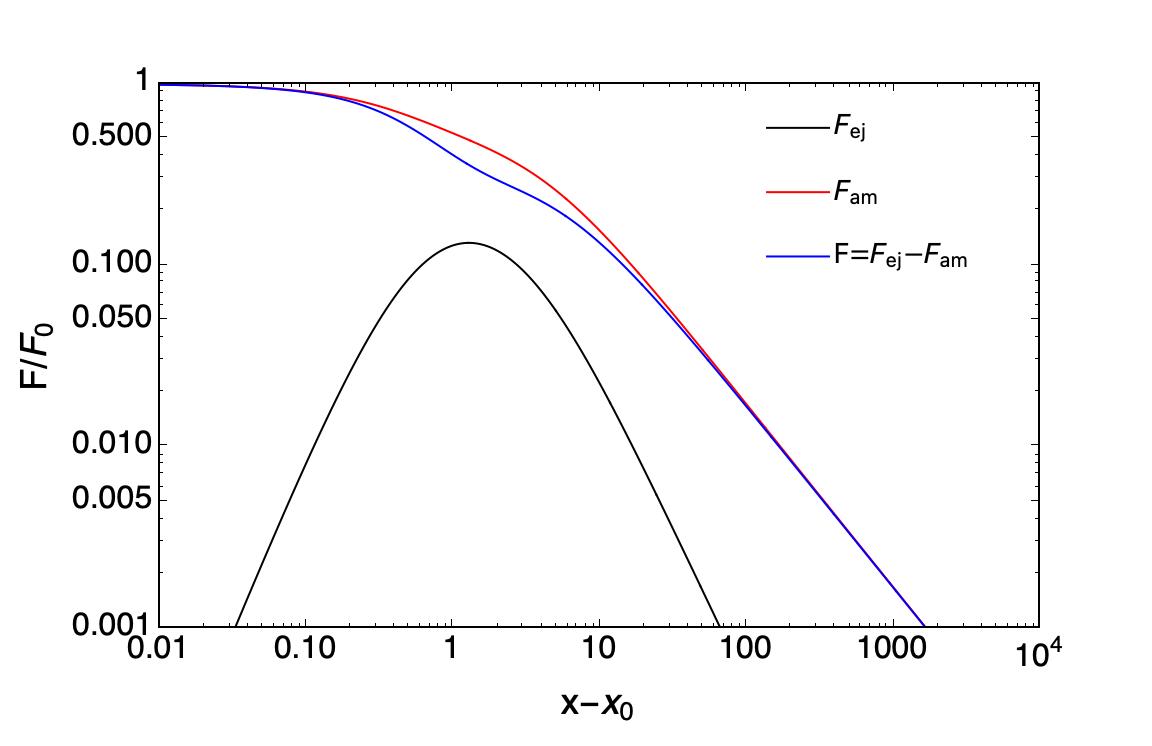}}
\caption{Time evolution of the expanding shell. Panels (a), (b), (c), (d), and (e) represent the time evolution of the shell radius, the shell mass, the mass addition rate to the shell, the force acting on the shell, respectively (see Section~\ref{sec:fidmodel} for the detailed formats). Note that the force curve is drawn by $|F|=|F_{\rm ej}-F_{\rm am}|$ because $F_{\rm am}>F_{\rm ej}$. We adopt the following set of parameters: $(n,\,s,\,\epsilon,\,\eta,\,\delta)=(5/3,\,1.0,\,1.0,\,1.0,\,1.0)$ as the fiducial model. 
}  
\label{fig:fidmodel}
\end{figure}

%
%%%%%%%%%%%%%%
\subsection{Fiducial model}
\label{sec:fidmodel}
%%%%%%%%%%%%%%
%

We make the fiducial model of the expanding shell. Here we adopt the following set of parameters: $(n,\,s,\,\epsilon,\,\eta,\,\delta)=(5/3,\,1.0,\,1.0,\,1.0,\,1.0)$ to examine the properties of the solution for the dimensionless equation of motion (see equation~\ref{eq:eom2}). The four panels of Figure~\ref{fig:fidmodel} depict how {each physical quantity} of the {expanding} thin shell {evolves} with time. Panel (a) shows the shell radius as a function of time.
{
The vertical and horizontal axes denote $y-y_0$ and $x-x_0$, respectively, where $y=R(t)/r_0$, $y_0=1$, $x=t/t_0$, and $x_0=1$. 
}
%The vertical and horizontal axes show the shell radius measured from $r_0$, and the time measured from $t_0$, respectively. Both axes are normalized by $r_0$ and $t_0$, and defined as $y=(R(t)-r_0)/r_0$ and $x=(t-t_0)/t_0$. 
The solid black and dashed red lines represent the numerical solution and the free expansion case, $y-y_0=(x-x_0)/\epsilon$, respectively. We find from the panel that the shell is decelerated at a very early time, $x-x_0\approx\mathcal{O}(1)$.

Panel (b) shows the time evolution of $M_{\rm ej}(t)$, $M_{\rm am}(t)$, $\Delta{m}$, and the sum of them (total mass). The normalized shell mass is given by
\begin{equation}
\frac{M(t)}{M_0}=\frac{1}{n-1}
\Biggr[
1-
\left[
x+\epsilon(1-y)
\right]^{1-n}
\Biggr]
+
\eta\epsilon
\frac{1}{3-s}
(y^{3-s}-1)
+
\epsilon\delta,
\label{eq:nond-totm}
\end{equation}
where $M_0\equiv\dot{M}_0\,t_0$. The solid black, red, blue, and dashed black lines denote $M_{\rm ej}(t)$, $M_{\rm am}(t)$, $\Delta{m}$, and $M(t)=M_{\rm ej}(t)+M_{\rm am}(t)+\Delta{m}$, respectively. The total mass approaches $M_{\rm am}(t)$ asymptotically.

Panel (c) depicts the rates at which ejecta and ambient matter are added to the shell as a function of time and also how the sum of them varies over time are shown. 
The time derivative of equation~(\ref{eq:nond-totm}) yields
\begin{equation}
\frac{\dot{M}(t)}{\dot{M}_0}=
(1-\epsilon\dot{y})
[x+\epsilon(1-{y})
]^{-n}
+
\eta\epsilon
y^{2-s}
\dot{y}.
\nonumber 
\label{eq:non-mdotej}
\end{equation}
It is noted from the panel that $\dot{M}_{\rm ej}(t)$ has a peak at the early time $x\sim1$, while $\dot{M}_{\rm am}(t)$ dominates over $\dot{M}_{\rm ej}(t)$ from early to late times. Thus, $\dot{M}(t)$ asymptotes to $\dot{M}_{\rm am}(t)$ at late times.

Panel (d) represents the evolution of the force acting on the shell over time. The normalized force is evaluated to be 
\begin{eqnarray}
\frac{F(t)}{F_{\rm ej,0}}=
\left[
x+\epsilon(1-y)\right]^{-n}
\left(
\frac{1}{\epsilon}-\dot{y}
\right)^2
-
\eta
y^{2-s}
\dot{y}^2,
\label{eq:netforce}
\end{eqnarray}
where we used equations~(\ref{eq:force})-(\ref{eq:eta0}) for the evaluation. The panel indicates that $F_{\rm ej}(t)$ has a peak in the early time $x\sim1$, affecting $F(t)$ there, and also $F_{\rm am}(t)$ dominates $F_{\rm ej}(t)$ from early to late times. Consequently, we find that $F(t)$ asymptotes to $F_{\rm am}(t)$ in late times.

%
%
%%%%%%%%%%%%%%%%%%%%%%%%%%%%%%%%%%%
\subsection{Power-law index of time on the shell radius evolution}
\label{sec:asymp-sol} 
%%%%%%%%%%%%%%%%%%%%%%%%%%%%%%%%%%%
%
%

Since $M(t)$ and $F(t)$ asymptote to $M_{\rm am}(t)$ and $F_{\rm am}(t)$, respectively, far enough away from the wind source, we can reevaluate $M(t)$ and $F(t)$ using equations~(\ref{eq:totm}) and (\ref{eq:force}) as
\begin{eqnarray}
M(t)
&&
\approx
\frac{4\pi\rho_{\rm am,0}r_0^3}{3-s}\left(\frac{R(t)}{r_0}\right)^{3-s},
\nonumber \\
F(t)
&&
\approx
-4\pi\rho_{\rm am,0}r_0^2
\left(\frac{R(t)}{r_0}\right)^{2-s}\dot{R}(t)^2,
\nonumber
\end{eqnarray}
respectively. Equation~(\ref{eq:eom}) is then reduced to be
\begin{eqnarray}
R(t)\ddot{R}(t)
=
(s-3)
\dot{R}(t)^2
\nonumber 
\label{eq:anas}
\end{eqnarray}
for ${R(t)}\gg{r_0}$. 
This second-order differential equation provides the simple power-law solution:
\begin{equation}
R(t)
\propto
t^{1/(4-s)}
\label{eq:sol}
\end{equation}
for $t\gg{t_0}$, which indicates that the shell evolves into a momentum-driven snow plow phase at a very late time.

\begin{figure}[ht!]
\centering
\includegraphics[scale=0.5]{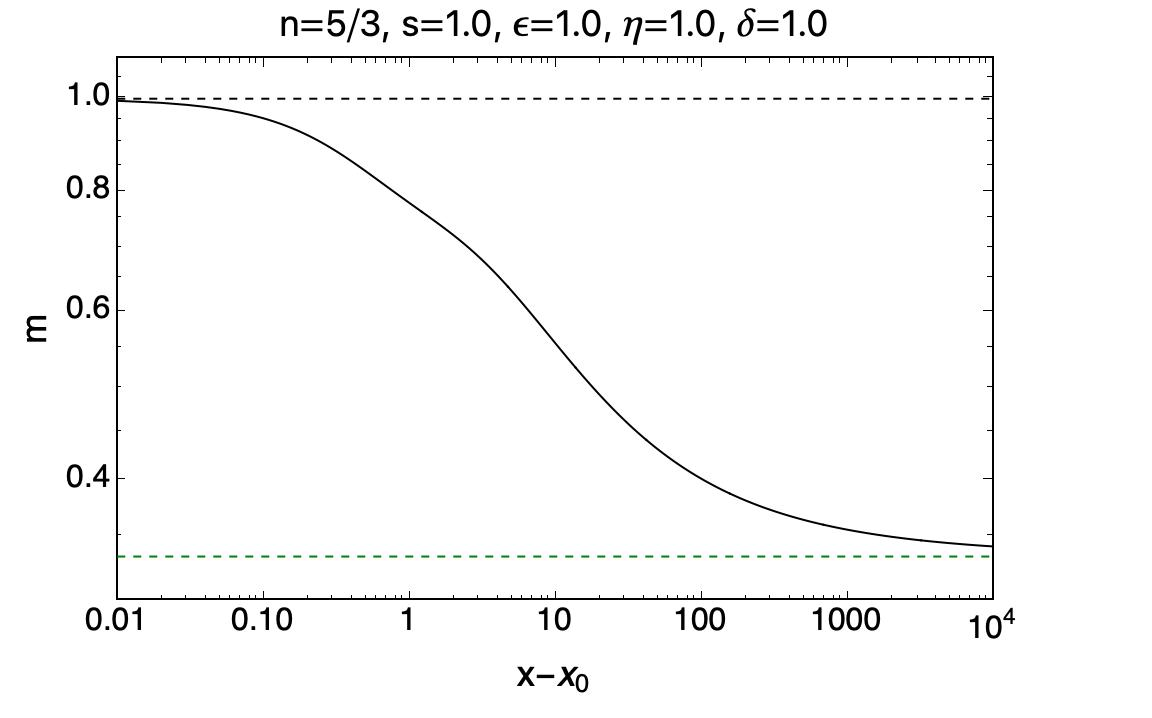}
\caption{
Time evolution of the power-law index of time, $m$, for the shell radius (see equation~\ref{eq:pl}) with the fiducial parameters: $(n,\,s,\,\epsilon,\,\eta,\,\delta)=(5/3,\,1.0,\,1.0,\,1.0,\,1.0)$. The dashed black and green lines represent the free expansion ($m=1$) and asymptotic solutions ($m=1/(4-s)$), respectively, whereas the solid black line denotes the evolution of $m$.
}
\label{fig:fidpl}
\end{figure}

The power-law index of time for the shell radius at the local time is defined by
\begin{eqnarray}
m\equiv
\frac{d\ln(y-y_0)}{d\ln(x-x_0)}.
%\nonumber 
\label{eq:pl}
\end{eqnarray}
The evolution of $m$ with the fiducial parameters: $(n,\,s,\,\epsilon,\,\eta,\,\delta)=(5/3,\,1.0,\,1.0,\,1.0,\,1.0)$ is depicted in Figure~{\ref{fig:fidpl}}. The solid black, dashed black, and dashed green lines represent the numerical solution, the free expansion ($m=1$) solution, and the asymptotic ($m=1/(4-s)$) solution, respectively. We find that $m$ decreases with time from $m=1$, indicating the shell decelerates and also its deceleration rate changes with time. We also confirm that the numerical solution approaches the asymptotic solution, which is given by equation~(\ref{eq:sol}), at $x\gg{x}_0$. Moreover, the figure demonstrates the significant variation of $m$ around $x-x_0=1$, which is caused by the concave on the $F(t)$ curve due to the ejecta's ram pressure, as shown in Figure~{\ref{fig:fidmodel}}(d).

%
%%%%%%%%%
% Figure
%%%%%%%%%
%

%\newpage

%
%%%%%%%%%%%%%%%%%%%%%%%%%%%%%
\subsection{Parameter-dependence on the shell evolution}
\label{sec:parameter-depend}
%%%%%%%%%%%%%%%%%%%%%%%%%%%%%
%

%
%%%%%%%%%
% Figure
%%%%%%%%%
%
\begin{figure}[ht!]
\centering
\subfigure[]{\includegraphics[scale = 0.44]{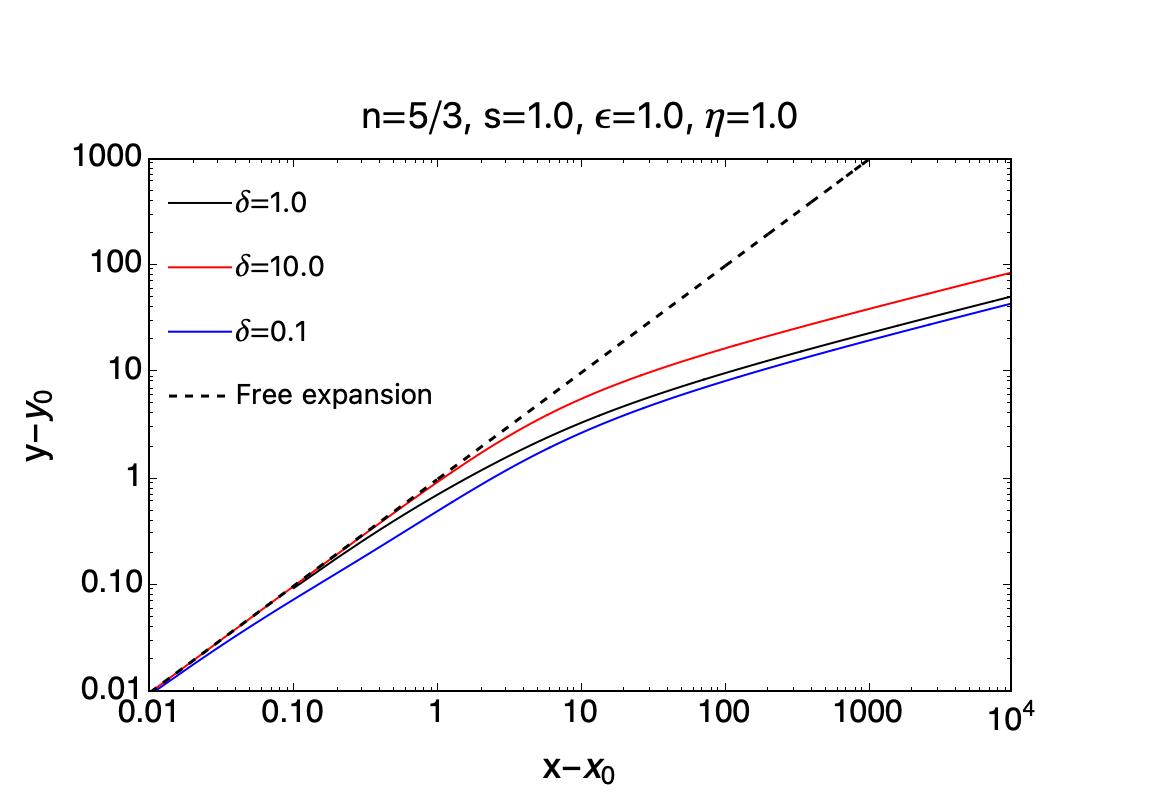}}
\subfigure[]{\includegraphics[scale = 0.44]{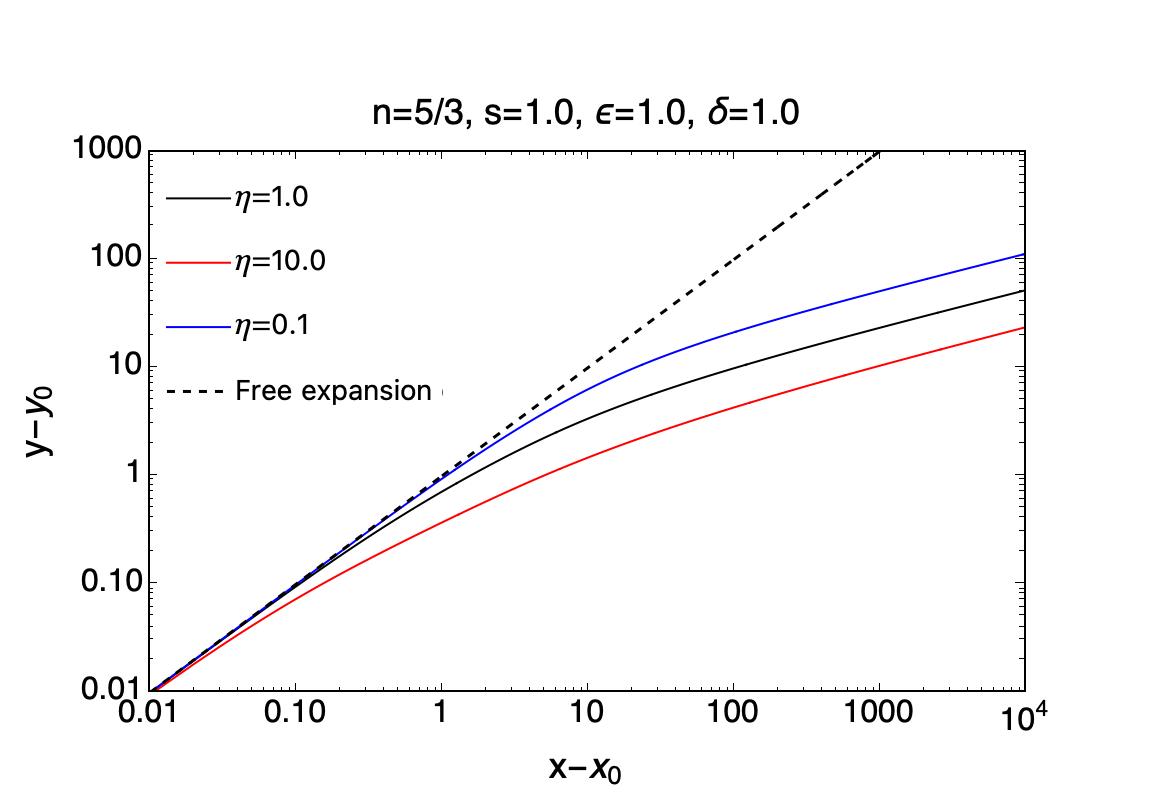}}
\subfigure[]{\includegraphics[scale = 0.44]{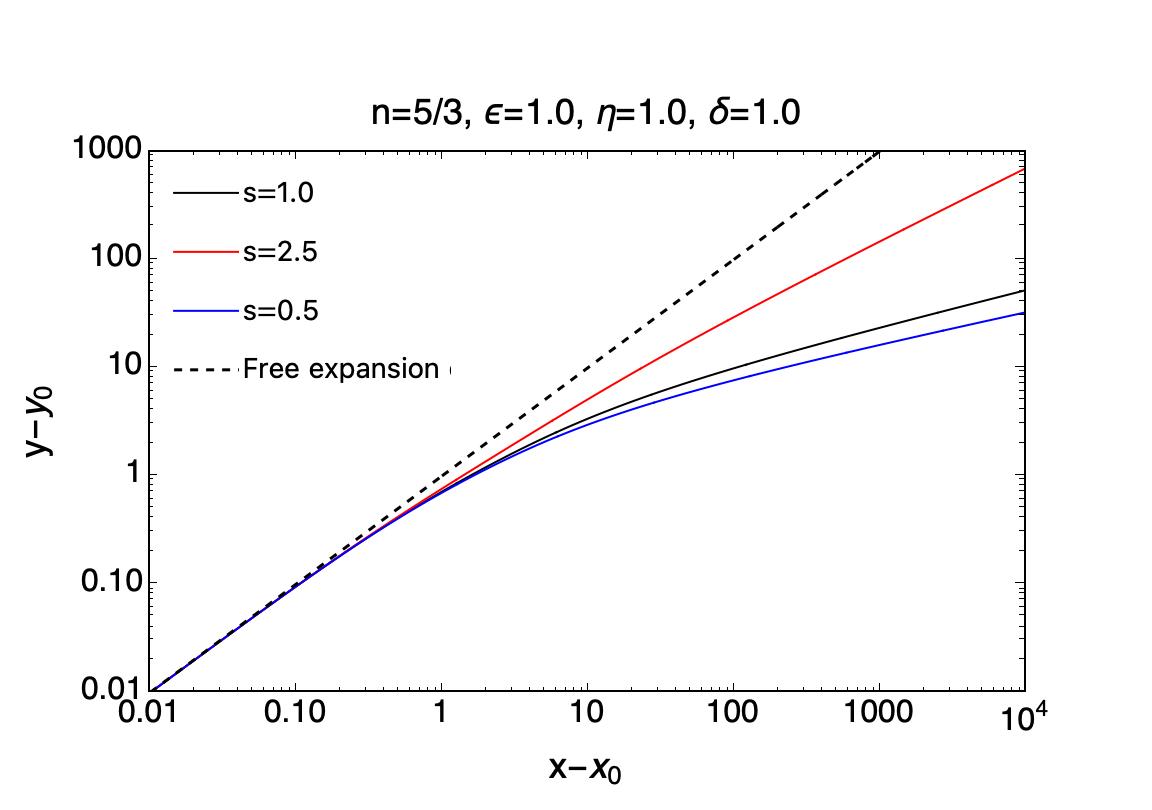}}
\subfigure[]{\includegraphics[scale = 0.44]{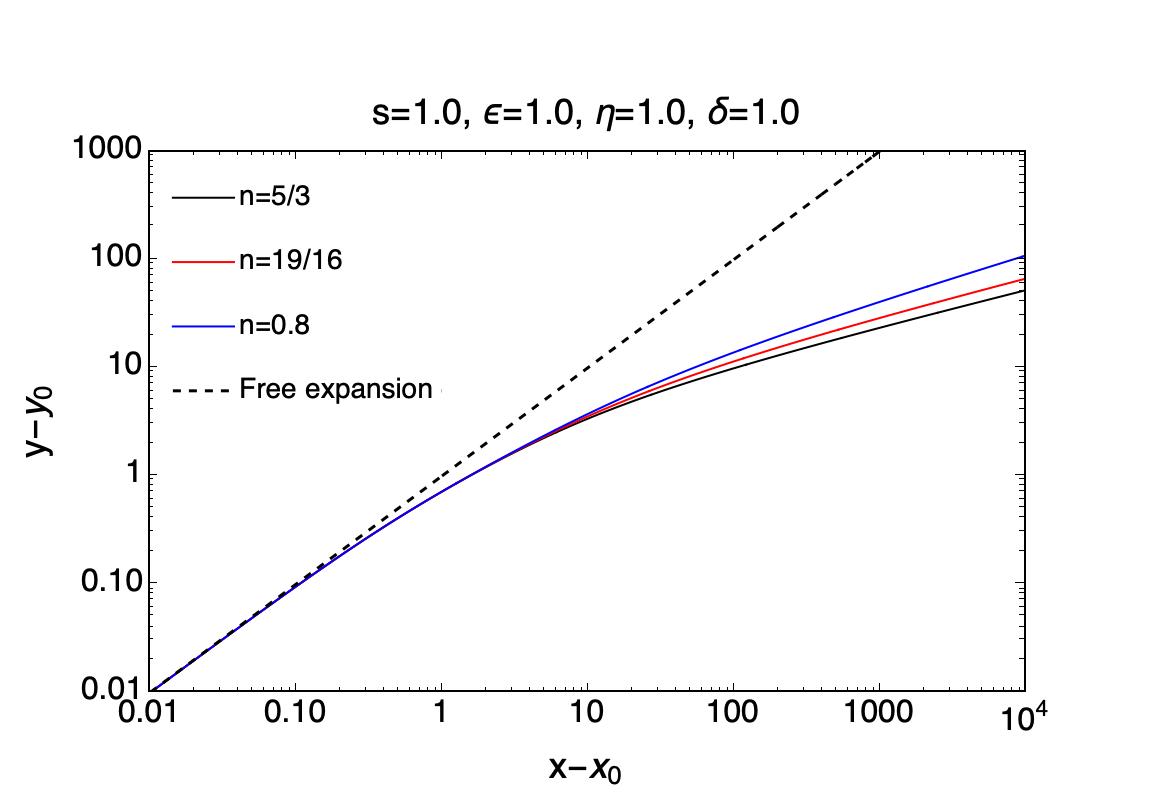}}
\subfigure[]{\includegraphics[scale = 0.44]{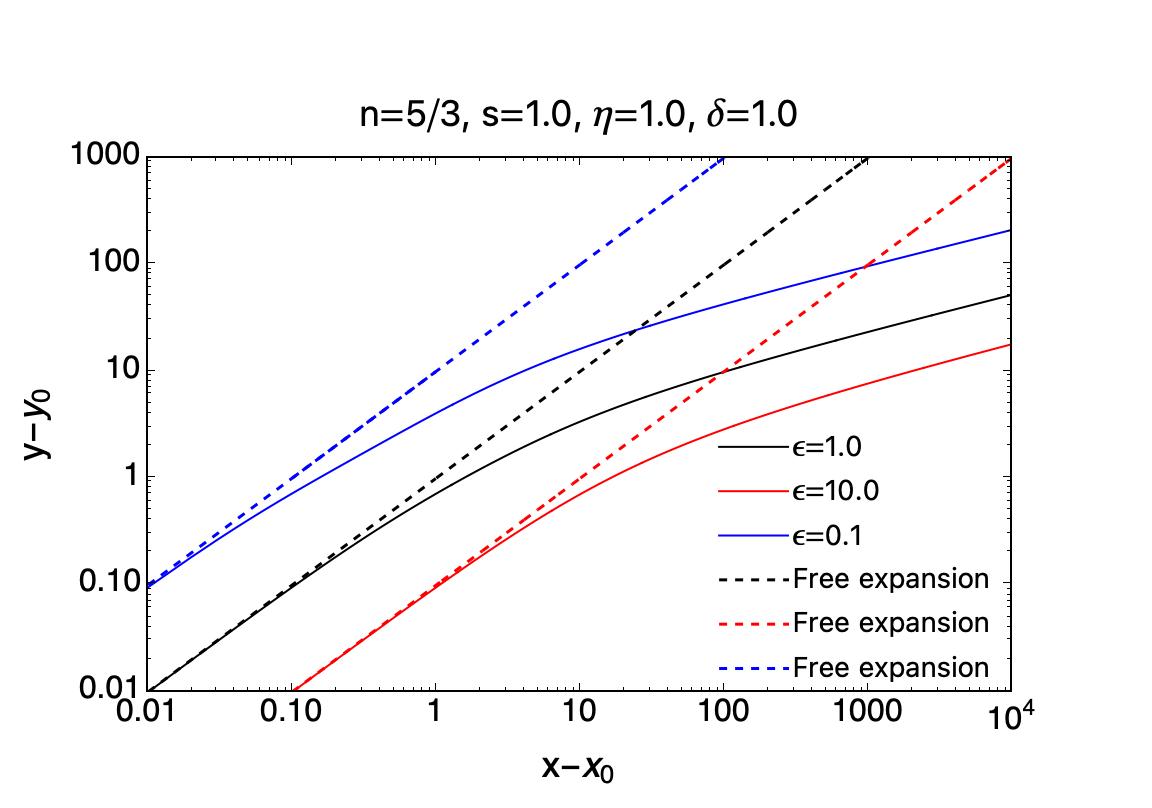}}
\caption{
Time evolution of the shell radius with five different panels.\,\,Panels (a) to (e) depict the dependence of the shell radius evolution on $\delta$, $\eta$, $s$, $n$, and $\epsilon$. In all panels, the dashed lines denote the shell radius evolving by free expansion, $y-y_0=(x-x_0)/\epsilon$. The different color represents the different value of each parameter.
}
\label{fig:shellradevo}
\end{figure}
%%%%%%
%
Next, we investigate how the time evolution of the expanding shell depends on the five parameters. Figure~\ref{fig:shellradevo} shows the shell radius as a function of time with the five different panels. In all panels, the dashed line indicates the shell radius evolving with free expansion velocity. Each panel represents the parameter dependence of the shell radius evolution and the deviation from the fiducial model.

In Figure~\ref{fig:shellradevo}(a), the solid black, red, and blue lines represent the numerical solutions with $\delta=1$, $\delta=10$, and $\delta=0.1$, respectively. We find that the shell decelerates with time and therefore the shell radius deviates from the free expansion case. The higher value of $\delta$ makes a slower deceleration than that of the lower value of $\delta$. By remembering $\delta$ controls the initial mass from equation~(\ref{eq:inimass}), we see this slower deceleration as the shell initially has a greater inertia.

In Figure~\ref{fig:shellradevo}(b), the solid black, red, and blue lines denote the numerical solutions with $\eta=1$, $\eta=10$, and $\eta=0.1$, respectively. According to equations~(\ref{eq:eta0}) and (\ref{eq:netforce}), the force acting against the shell expansion, $F_{\rm am}(t)$, gets stronger as the value of $\eta$ is larger. That is why the deceleration rate of the $\eta=10$ case is higher than the other two cases.

In Figure~\ref{fig:shellradevo}(c), the solid black, red, and blue lines depict the numerical solutions with $s=1$, $s=2.5$, and $s=0.5$, respectively. Equation~(\ref{eq:rhoam}) shows that the ambient matter density rapidly decreases with radius as the value of $s$ is larger, suggesting that the larger value of $s$ makes the deceleration rate lower. This is also demonstrated by equation~(\ref{eq:netforce}), which indicates that the force acting against the shell expansion gets weaker as the value of $s$ is larger. 

In Figure~\ref{fig:shellradevo}(d), the solid black, red, and blue lines represent the numerical solutions with $n=5/3$, $n=19/16$, and $n=0.8$, respectively. We choose these values because $n=5/3$ is the standard value of the mass fallback rate \citep{evans_tidal_1989}, and $n=19/16$ and $n=0.8$ are obtained based on the self-similar solutions of a TDE disk \citep{cannizzo_disk_1990, mummery_spectral_2020}. Equation~(\ref{eq:rhoej}) indicates that the ejecta density more slowly decreases with time as the value of $n$ is lower, implying that it makes the deceleration rate lower. This is also seen by equation~(\ref{eq:netforce}), which states that the force acting on expanding the shell gets stronger for the lower value of $n$.

In Figure~\ref{fig:shellradevo}(e), the solid black, red, and blue lines display the numerical solutions with $\epsilon=1$, $\epsilon=10$, and $\epsilon=0.1$, respectively. The initial velocity is higher as the value of $\epsilon$ is lower. In other words, in comparison with the fiducial model of $\eta=1$, the shell radius is more delayed in expanding in the $\epsilon=10$ case, while it expands faster in the $\epsilon=0.1$ case. 

%
%%%%%%%%%
% Figure
%%%%%%%%%
%
\begin{figure}[ht!]
\centering
\subfigure[]{\includegraphics[scale = 0.44]{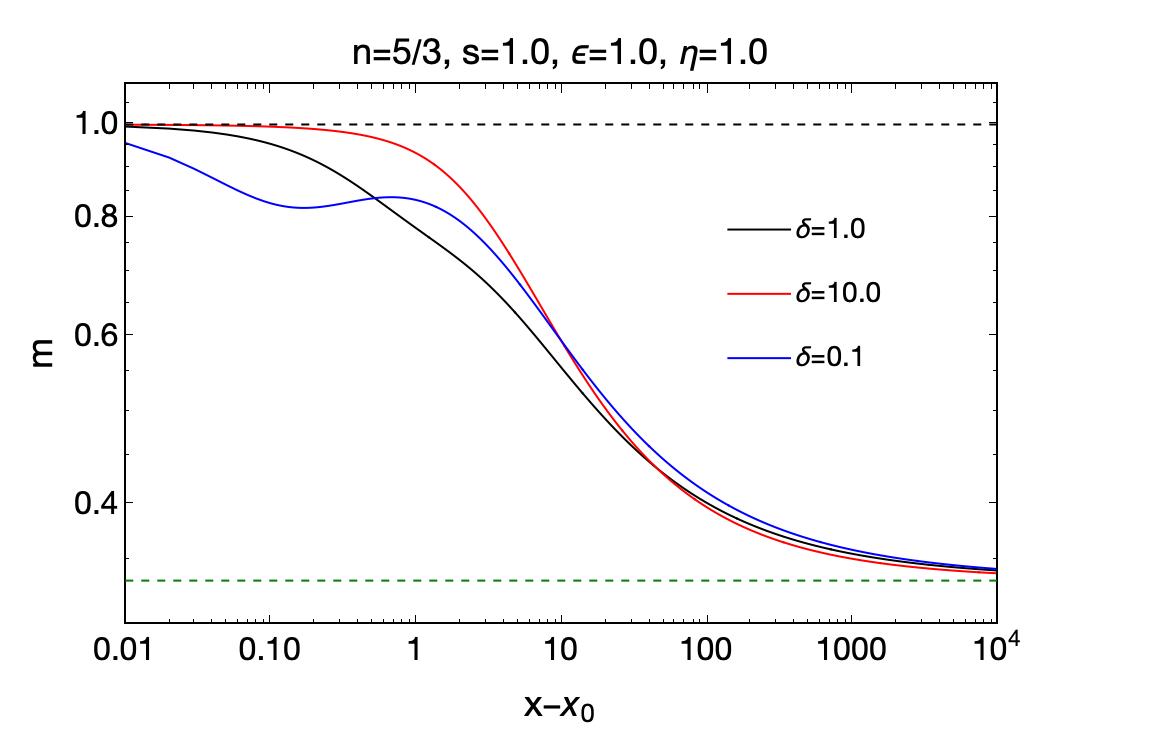}}
\subfigure[]{\includegraphics[scale = 0.44]{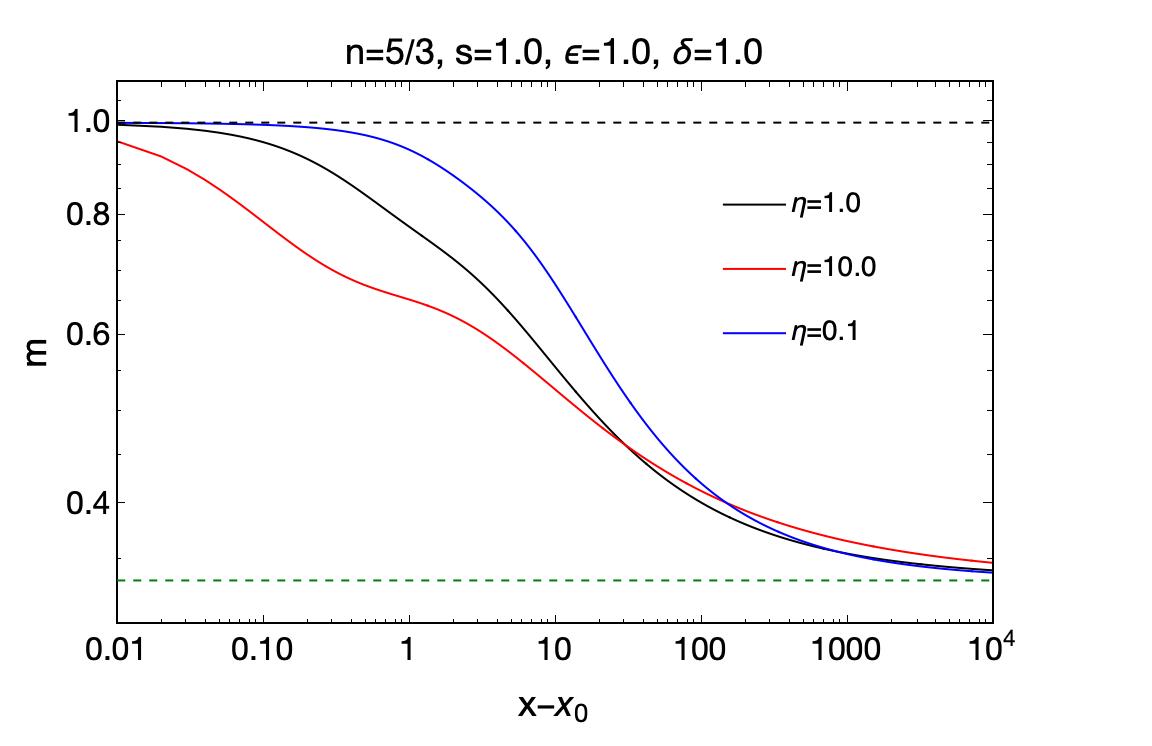}}
\subfigure[]{\includegraphics[scale = 0.44]{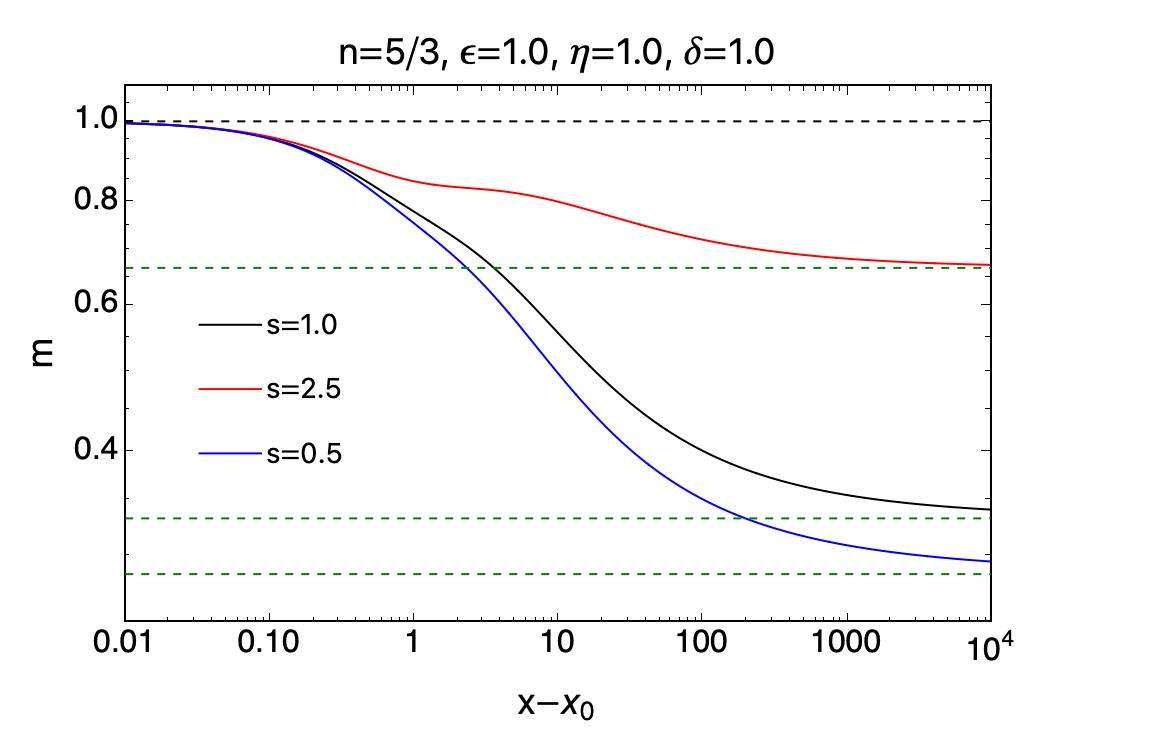}}
\subfigure[]{\includegraphics[scale = 0.44]{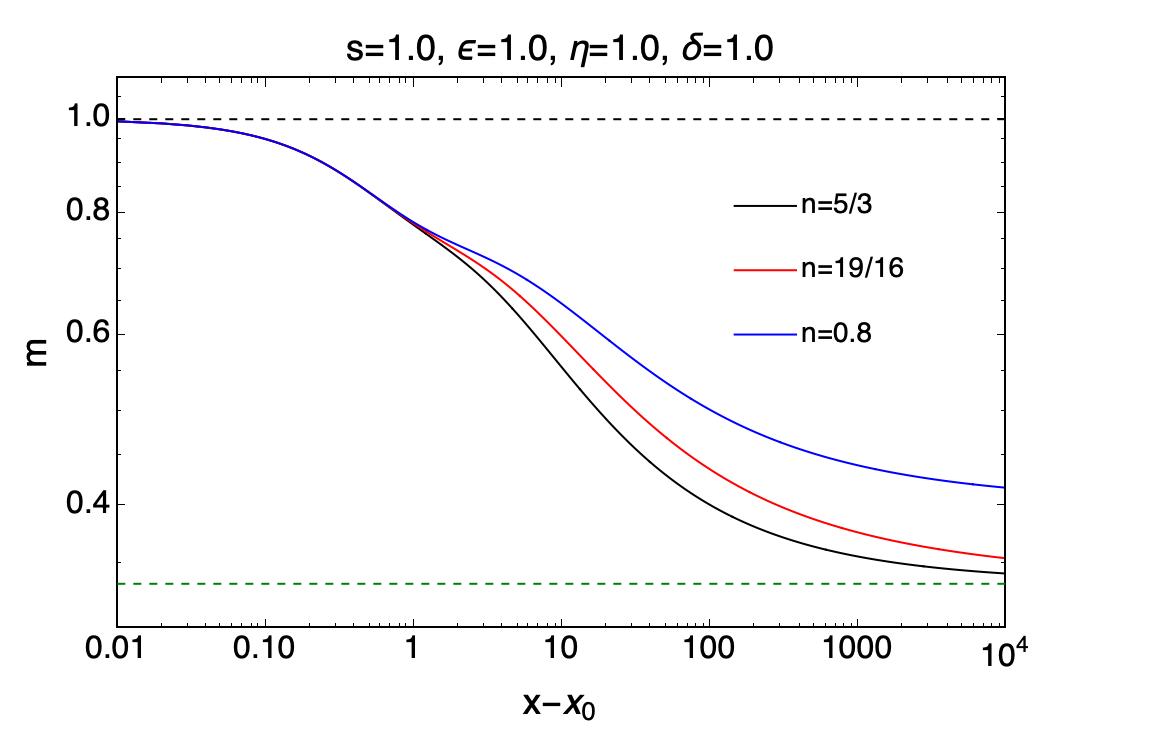}}
\subfigure[]{\includegraphics[scale = 0.44]{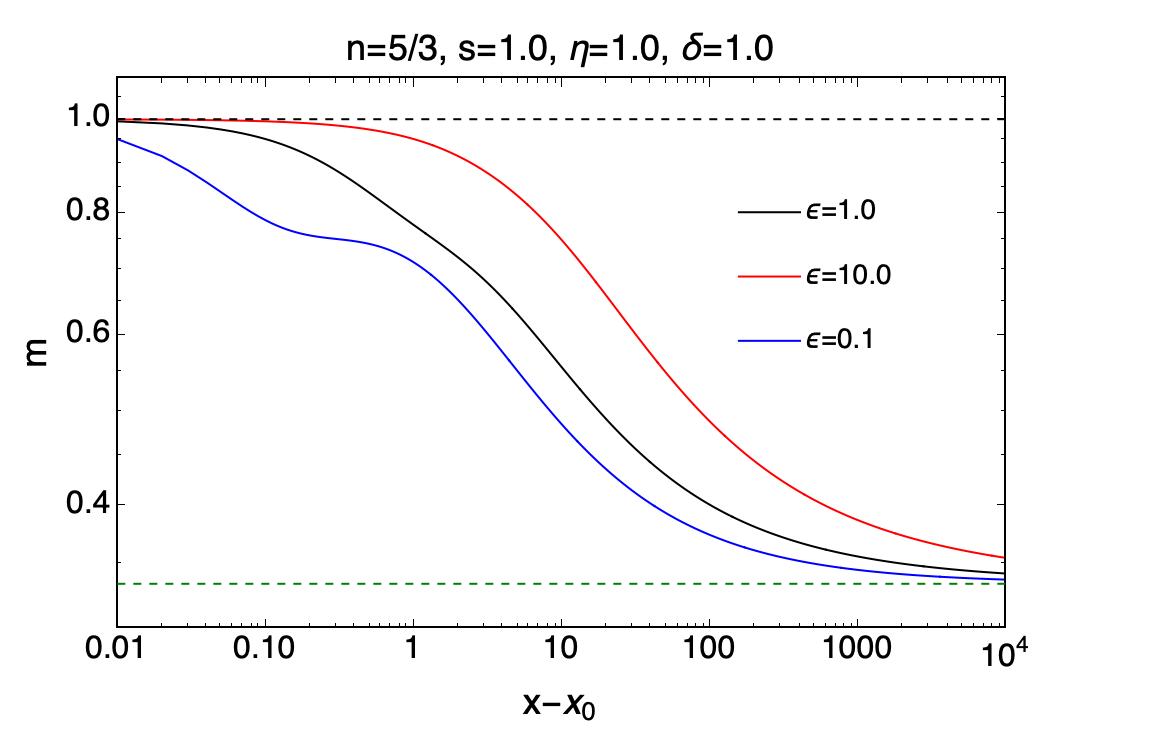}}
\caption{
Time evolution of the local power-law index of time for the shell radius with five different panels. Panels (a) to (e) represent the dependence of the power-law index evolution on $\delta$, $\eta$, $s$, $n$, and $\epsilon$. The different color shows the different value of each parameter.
}
\label{fig:plevo}
\end{figure}

Figure~\ref{fig:plevo} shows the time evolution of the local power-law index of time, $m$, for the shell radius with the five different panels. Each panel illustrates the parameter-dependence of the power-law index and the deviation from the fiducial model. In all the panels, the black and green dashed lines denote the $m=1$ and $m=1/(4-s)$ lines, respectively. The former indicates the free expansion solution, while the latter represents the simple power-law solution at very late times, which is given by equation~(\ref{eq:sol}). Overall, the shell radius evolution follows no simple power-law of the time from early to middle/late times, but the local power-law index asymptotes to $m=1/(4-s)$ line at a very late time.

In Figure~\ref{fig:plevo}(a), the solid black, red, and blue lines represent the local power-law index of time for the shell radius with $\delta=1$, $\delta=10$, and $\delta=0.1$, respectively. For all the values of $\delta$, $m$ decays from nearly $1.0$ to $1/3$ at a very late time. Also, a higher value of $\delta$ tends to take along the $m=1$ line at early times. This is because momentum inertia works more effectively for a more massive initial mass, as already given in the description of Figure~\ref{fig:shellradevo}(a). Notably, for the lightest initial mass (i.e. the $\delta=0.1$ case), the $m$ curve has a remarkable variation at early times. This is because the force acting on the shell from the ejecta peaks around $x/x_0\sim1$, significantly impacting the shell radius evolution.

In Figure~\ref{fig:plevo}(b), the solid black, red, and blue lines display the numerical solutions with $\eta=1$, $\eta=10$, and $\eta=0.1$, respectively. According to equation~(\ref{eq:netforce}), $F_{\rm am}(t)$ is weaker as the value of $\eta$ is lower, making the deceleration rate lower. For the $\eta=0.1$ case, the local power-law index for time takes along the $m=1$ line in early times. In contrast, the shell decelerates most effectively in the $\eta=10$ case and is more easily affected at early times by the ejecta consequently.

In Figure~\ref{fig:plevo}(c), the solid black, red, and blue lines depict the numerical solutions with $s=1$, $s=2.5$, and $s=0.5$, respectively. As $s$ is higher, the force due to the ambient matter is weaker, making the deceleration rate lower and thus approaching the free expansion solution. The three curves overlap at early times but start deviating around $x-x_0=0.1$. These tendencies are consistent with Figure~\ref{fig:shellradevo}(c).

In Figure~\ref{fig:plevo}(d), the solid black, red, and blue lines indicate the numerical solutions with $n=5/3$, $n=19/16$, and $n=0.8$, respectively. The three curves overlap at early times but start deviating around $x-x_0=1.0$. For the higher value of $n$, the local power-law index more rapidly decreases with time. These tendencies are consistent with Figure~\ref{fig:shellradevo}(d).

In Figure~\ref{fig:plevo}(e), the solid black, red, and blue lines denote the numerical solutions with $\epsilon=1$, $\epsilon=10$, and $\epsilon=0.1$, respectively. It is noted from equation~(\ref{eq:netforce}) that as the value of $\epsilon$ is higher, the net force is weaker. This suggests that the shell is harder to decelerate for the higher value of $\epsilon$ so that the local power-law index stays at $m=1$ in early times for the $\epsilon=10$ case. In contrast, the $\epsilon=0.1$ case illustrates that the local power-law index deviates initially from $m=1$, indicating the shell rapidly decelerates from the beginning because the net force is strongest among the three cases. That is why $m$ varies with time significantly.

%
%%%%%
\newpage
%%%%%
%

%
%%%%%%%%%%%%%%%%%%%%%%%%%%%%%%
\section{Application to radio observations of AT2019dsg}
\label{sec:obs}
%%%%%%%%%%%%%%%%%%%%%%%%%%%%%%
%
\cite{stein_tidal_2021} first made the spectral analysis of radio emissions from AT2019dsg, and subsequently, \cite{cendes_radio_2021} examined them in more detail and longer term. Both of them obtained the synchrotron peak flux $F_{\rm p}$ and corresponding frequency $\nu_{\rm p}$ by fitting the observed flux to a synchrotron self-absorbed spectral model \citep{chevalier_synchrotron_1998,barniol_duran_radius_2013}. They evaluated various quantities at different epochs, including the size of the radio-emitting region $R_{\rm eq}$ and the ambient number density $n_{\rm ext}$, through the equipartition analysis with $\nu_{\rm p}$ and $F_{\rm p}$. Table 2 of \citep{cendes_radio_2021} summarizes the detailed numerical values of measured quantities. The resultant $R_{\rm eq}$ very weakly depends on $\nu_{\rm p}$ and $F_{\rm p}$, indicating that $R_{\rm eq}$ is robustly determined. We assume the expanding shell radius equals the size of the radio-emitting region, enabling a direct comparison of our model and $R_{\rm eq}$.

%
%%%%%%%%%%%%%%%%%%%%%%%%%%%
\subsection{Identification of normalization parameters}
%%%%%%%%%%%%%%%%%%%%%%%%%%%
%

We evaluate the normalization parameters of our numerical model by applying the observed values of TDE candidate AT2019dsg \citep{cendes_radio_2021}. We adopt the black hole mass as $10^{6.7}\, M_\odot$ and a solar-type star for the stellar mass and radius.
The normalization parameters are then evaluated as follows: 
\begin{eqnarray}
&&
r_0=r_{\rm t}=\left(\frac{M}{m_*}\right)^{1/3}r_*
\sim1.1\times10^{13}\,{\rm cm}\,\left(\frac{M}{10^{6.7}\,M_\odot}\right)^{1/3}
\left(\frac{m_*}{M_\odot}\right)^{-1/3}
\left(\frac{r_*}{R_\odot}\right),
\label{eq:r0}
\\
&&
t_0
=\sqrt{\frac{r_*^3}{Gm_*}}
\sim1.6\times10^{3}\,{\rm s}\,
\left(\frac{m_*}{M_\odot}\right)^{-1/2}
\left(\frac{r_*}{R_\odot}\right)^{3/2},
\label{eq:t0}
\\
&&
\frac{V_0}{c}
=\sqrt{\frac{GM}{c^2r_{\rm t}}}=\left(\frac{M}{m_*}\right)^{1/3}\left(\frac{Gm_*}{r_*}\right)^{1/2}
\sim0.25
\left(\frac{M}{10^{6.7}\,M_\odot}\right)^{1/3}
\left(\frac{m_*}{M_\odot}\right)^{1/6}
\left(\frac{r_*}{R_\odot}\right)^{-1/2},
\label{eq:v00}
\\
&&
M_0=\dot{M}_0t_0=
\frac{f}{3}m_*
\left(
\frac{t_0}{t_{\rm mtb}}
\right)
%=\frac{\sqrt{2}}{3\pi}m_*\left(\frac{m_{*}}{M}\right)^{1/2}
\simeq
6.7\times10^{-6}
\,M_\odot\,
\left(\frac{f}{0.1}\right)
\left(\frac{M}{10^{6.7}\,M_\odot}\right)^{-1/2}
\left(\frac{m_*}{M_\odot}\right)^{3/2},
\label{eq:m0}
\end{eqnarray}
where $\dot{M}_0=(f/3)(m_*/t_{\rm mtb})$, $f$ is the fraction of mass loss to fallback or accretion rates; its typical value is $f=0.1$ \citep{strubbe_optical_2009}, and the orbital period of the stellar debris on the most tightly bound orbit is given by
\begin{equation}
t_{\rm mtb}=\frac{\pi}{\sqrt{2}}\left(\frac{M}{m_*}\right)^{1/2}t_0
\sim92\,{\rm day}\,
\left(\frac{M}{10^{6.7}\,M_\odot}\right)^{1/2}
\left(\frac{m_*}{M_\odot}\right)^{-1}
\left(\frac{r_*}{R_\odot}\right)^{3/2}.
\nonumber
\label{eq:mtb}
\end{equation}
Note that $\dot{M}_0=(f/3)(m_*/t_{\rm mtb})$ we adopted here implies the disk formation timescale is significantly shorter than $t_{\rm mtb}$. If the disk formation timescale is longer than $t_{\rm mtb}$, $\dot{M}_0$ is smaller than the current case (see Section 2 of \citealt{hayasaki_origin_2021} for the detail).

The initial density of the ejecta is evaluated to be
\begin{eqnarray}
\rho_{\rm ej,0}
&&
=\frac{\dot{M}_0}{4\pi{r_0^2}v_0}
\sim
%1.26
1.3
\times
10^{-12}
\,{\rm g/cm^3}\,
\left(\frac{f}{0.1}\right)
\left(\frac{\epsilon}{1.0}\right)
\left(\frac{M}{10^{6.7}\,M_\odot}\right)^{-3/2}
\left(\frac{m_*}{M_\odot}\right)^{5/2}
\left(\frac{r_*}{R_\odot}\right)^{-3}.
\label{eq:rhoej00}
\end{eqnarray}
Using equations~(\ref{eq:rhoam}) and (\ref{eq:r0}), we evaluate the initial density of the ambient matter as
\begin{equation}
\rho_{\rm am,0}=\rho_{\rm obs}\left(\frac{r}{r_0}\right)^{s}
\sim
1.7
\times10^{-14}\,{\rm g/cm^3}
\left(\frac{n_{\rm ext}}{10^{3.56}\,{\rm cm^{-3}}}\right)
\left(\frac{r}{10^{16.15}\,{\rm cm}}\right)^{2.1}
\left(
\frac{r_{0}}
{
r_{\rm t}
}
\right)^{-2.1},
\label{eq:rhoam0}
\end{equation}
where $\rho_{\rm obs}\simeq{m}_{\rm p}\,n_{\rm ext}$ with the proton mass $m_{\rm p}$, and $n_{\rm ext}=10^{3.56}\,{\rm cm^{-3}}$ at $r=10^{16.15}\,{\rm cm}$ and $s=2.1$ are adopted as fiducial values, respectively \citep{cendes_radio_2021}.
Substituting equations~(\ref{eq:rhoej00}) and (\ref{eq:rhoam0}) into (\ref{eq:eta0}) yields the relation between $\eta$ and $\epsilon$ as
\begin{eqnarray}
\eta\,\epsilon
&&
=
\frac{4\pi{r_0^3}\rho_{\rm obs}}{\dot{M}_0\,t_0}
\left(\frac{r}{r_0}\right)^s
\nonumber \\
&&
\sim
2.8\times
10^{-2}
\left(\frac{f}{0.1}\right)^{-1}
\left(\frac{n_{\rm ext}}{10^{3.56}\,{\rm cm^{-3}}}\right)
\left(\frac{r}{10^{16.15}\,{\rm cm}}\right)^{2.1}
\left(
\frac{r_{0}}
{
r_{\rm t}
}
\right)^{0.9}
\left(\frac{M}{10^{6.7}\,M_\odot}\right)^{1/2}
\left(\frac{m_*}{M_\odot}\right)^{-3/2}
\label{eq:etaepsilon}
\end{eqnarray}
The initial shell mass $\Delta{m}$ is evaluated to be
\begin{eqnarray}
\Delta{m}
=\delta\epsilon\dot{M}_0t_0
\simeq
6.7\times
10^{-6}
\,M_\odot\,
\left(\frac{f}{0.1}\right)
\left(\frac{\delta}{1.0}\right)
\left(\frac{\epsilon}{1.0}\right)
\left(\frac{M}{10^{6.7}\,M_\odot}\right)^{-1/2}
\left(\frac{m_*}{M_\odot}\right)^{3/2}.
\label{eq:delta}
\end{eqnarray}
This value of the initial shell mass is reasonable for the shell formation due to the TDE disk wind because the initial shell mass is much smaller than the fallback mass for the reasonable range of value of $\delta$ and $\epsilon$. 

%
%
%%%%%%%%%%%%%%%%%%%%%
\subsection{Comparison to observations}
\label{sec:compobs}
%%%%%%%%%%%%%%%%%%%%%
%
For direct comparison purposes with the observed radio-emitting radius $R_{\rm eq}$ and ambient matter density $\rho_{\rm obs}\simeq{m_{\rm p}}n_{\rm ext}$ of AT2019dsg (see Table 2 of \citealt{cendes_radio_2021}), we need to decide the five parameters and also match the time origin between the observed data and our model by introducing $T_{\rm d}$, which is defined as the onset time of OUV emissions (see also Table~\ref{tbl1} with the other epochs).

Once we decide those parameters, we obtain the dimensionless shell radius, $y(x)$, where $y = R(t)/r_0$ and $x=t/t_0$ (see also equation~\ref{eq:dlq}) by solving equation~(\ref{eq:eom2}) with them numerically. The physical value of $R(t)$ as a function of time is obtained by using equations (\ref{eq:r0}) and (\ref{eq:t0}).

Following the standard TDE scenario, we set $n=5/3$ as the fiducial value. We adopt $s=2.1$ because the power-law index of the ambient matter (the circumnuclear density noted in their paper) is $2.1$ on average according to \cite{cendes_radio_2021}. We can constrain $\epsilon$, which controls the initial velocity, as $0.25\lesssim\epsilon\lesssim3.5$ because the upper limit is the speed of light, and the lower limit is given by the observed value, $0.07\,c$. Once we adopt the fiducial vale of $\epsilon$ (i.e. $\epsilon=1.0$) for the current case, we obtain $\eta\sim2.8\times10^{-2}$ from equation~(\ref{eq:etaepsilon}). Regarding the remaining parameter $\delta$, we choose $\delta=5$ to fit in the observational data. Note that $\delta$ is such a parameter to set that $\Delta{m}$ is significantly smaller than the fallback mass (see equation~\ref{eq:delta}). The parameter set used for our numerical model is summarized as $(n,\,s,\,\epsilon,\,\eta,\,\delta)=(5/3,\,2.1,\,1.0,\,0.028,\,5.0)$.

\begin{figure}[ht!]
\centering
\subfigure[]{\includegraphics[scale = 0.58]{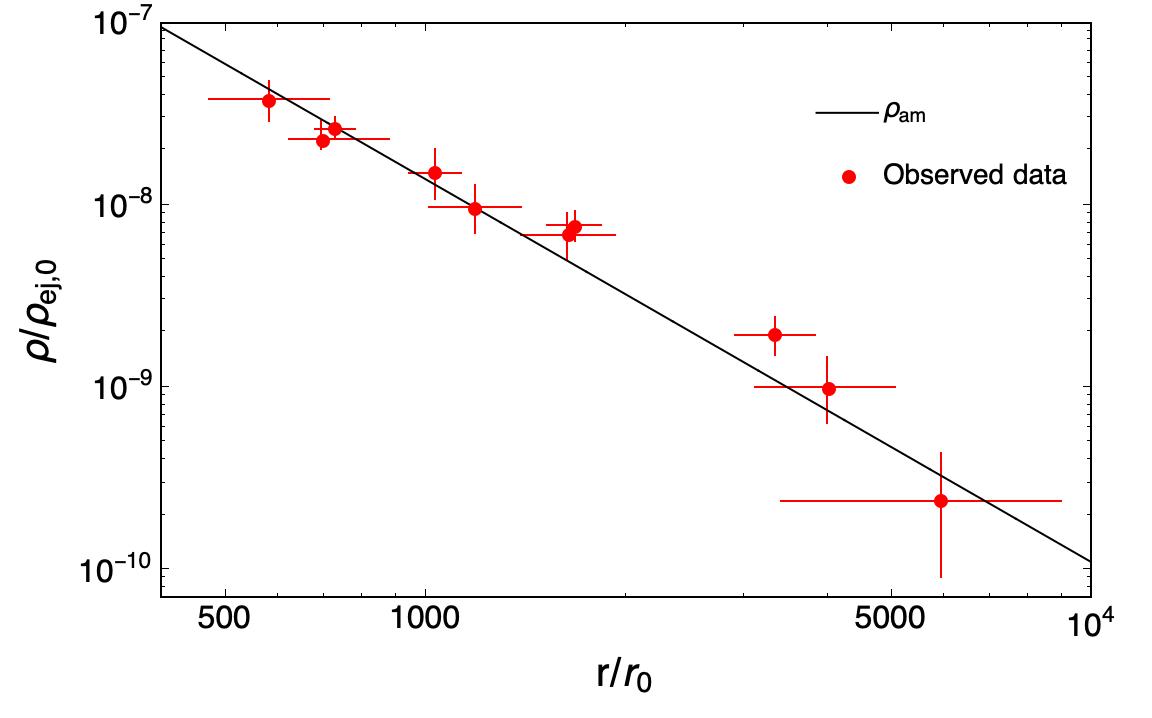}}
\subfigure[]{\includegraphics[scale = 0.58]{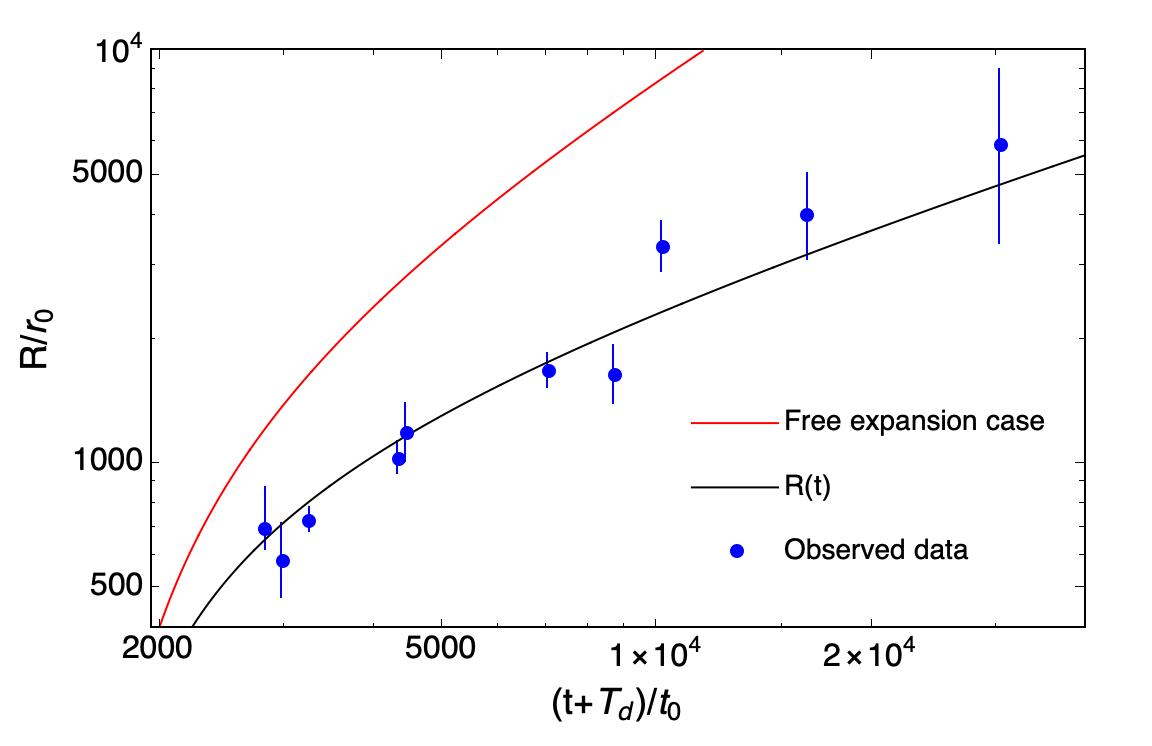}}
\caption{
Comparison of the radial profile of the ambient matter density (panel a) and the shell radius evolution (panel b) between our model and the observed data of AT2019dsg \citep{cendes_radio_2021}. In panel (a), the red-filled small circles with error bars and the solid black line depict, respectively, the observed data and the radial density profile, which is given by equation~(\ref{eq:rhoam}). The vertical and horizontal axes indicate the ambient matter density normalized by $\rho_{\rm ej,0}$ and the radius normalized by $r_0$, respectively. In panel (b), the solid black and red lines denote the free expansion and numerical solutions, respectively. The blue-filled small circles with the error bars depict the observed data. The vertical and horizontal axes denote $R(t)/r_0$ and $(t+T_{\rm d})/t_0$, respectively, where $T_{\rm d}$ is the onset time of OUV emissions (see also equation~\ref{eq:td} and Table~\ref{tbl1}). The following set of parameters: $(n,\,s,\,\epsilon,\,\eta,\,\delta)=(5/3,\,2.1,\,2.0,\,0.028,\,5.0)$ is adopted for the numerical solution, $R(t)$.
 }
\label{fig:r-t-obs}
\end{figure}

Two panels of Figure~\ref{fig:r-t-obs} display the comparison between our model and the observed data of AT2019dsg. Panel (a) depicts the radial profile of the ambient matter density. The red-filled small circles with error bars indicates the observed radial density profile $\rho_{\rm obs}(r)$, while the solid black line denotes our numerical solution, $\rho_{\rm am}(r)$. The vertical and horizontal axes represent the ambient matter density normalized by $\rho_{\rm ej,0}$ and the radius normalized by $r_0$, respectively. The panel demonstrates that our assumption about the ambient mass density distribution (see equation~{\ref{eq:rhoam}}) is consistent with the observed data.

Next, panel (b) illustrates how the shell radius varies over time, where the solid red line, solid black line, and blue-filled circles with error bars represent the free expansion and numerical solutions, and observational data, respectively. The vertical axis shows the shell radius normalized by $r_0$, while the horizontal axis shows $(t+T_{\rm d})/t_0$.
%\blue{These epochs are summarized in Table~\ref{tbl1} for clarification purposes.} 
Assuming $\dot{R}\sim{v_0}$ at $R=R_{\rm obs}$ and $T_{\rm d}\gg{t_0}$, $T_{\rm d}$ is approximately evaluated to be
\begin{eqnarray}
T_{\rm d}\lesssim(P_{\rm obs}-R_{\rm eq}/v_0),
\label{eq:td}
\end{eqnarray}
where we adopt $R_{\rm eq}=10^{15.84}\,{\rm cm}$ at $P_{\rm obs}=52\,{\rm day}$. It is noted from equations~(\ref{eq:epsilon0}), (\ref{eq:v00}), and (\ref{eq:td}) that {$T_{\rm d}\lesssim41\,{\rm days}$ for $\epsilon=1$}. The panel shows that the numerical solution is in good agreement with the observed data within the error bars. Overall, Figure~\ref{fig:r-t-obs} indicates that our model well reproduces the observed properties with reasonable model parameters.

%
%%%%%%%%%
% Figure
%%%%%%%%%
%

%
%%%%%%
% Table
%%%%%%
%
\begin{table}
\begin{center}
\begin{tabular}{lll}
%\tableline
%\tableline
%Epoch & Event name &\\
\tableline
$t=-T_{\rm d}$ & {\rm Onset of OUV emissions}\\
$t=0$ & {\rm Time origin of our model}\\
$t=t_0$ & {\rm Shell formation}\\
\tableline
\end{tabular}
\caption{Summary of events in our reference frame. They are listed in chronological order from top to bottom.\label{tbl1}}
\end{center}
\end{table}

%
%%%%%%%%%%
\section{Discussion}
\label{sec:disc}
%%%%%%%%%%
%
%
We have derived the dynamics of the expanding thin shell driven by mass loss from the accretion disk in TDEs. For the case of AT2019dsg, the ratio of the ambient ram to thermal pressures at the observed radius $R\sim10^{16}\,{\rm cm}$ is evaluated as
\begin{eqnarray}
\frac{P_{\rm am}}{P_{\rm th}}\sim0.9
\left(\frac{E_0}{10^{48}\,{\rm erg}}\right)^{-1}
\left(\frac{R}{10^{16}\,{\rm cm}}\right)^{3}
\left(\frac{n_{\rm am}}{10^4\,{\rm cm^{-3}}}\right)
\left(\frac{\dot{R}}{0.1c}\right)^2,
\nonumber
\end{eqnarray}
where the thermal pressure is estimated to be $P\sim(E_0/(4\pi{R}^3/3))$ if the radiation loss is negligibly small, 
$E_0\sim10^{48}\,{\rm erg}$ is the initially injected energy to the shell, $P_{\rm am}\sim{m_{\rm p}}{n}_{\rm am}\dot{R}^2$ is the ambient ram pressure with $n_{\rm am}\sim10^{4}\,{\rm cm^{-3}}$ and $\dot{R}\sim0.1c$ from the radio observations \citep{cendes_radio_2021}.
For the late time, the thermal pressure is comparable to or more than the ambient ram pressure and thus may affect the shell dynamics, although the simple interpretation tells us thermal pressure is just likely to act to weaken the deceleration of the expanding shell. Furthermore, incorporating thermal energy is crucial for the energy evolution within the expanding shell, a factor that can intrinsically connect to the energy injection problem regarding the radio emission of AT2019dsg (e.g. \citealt{matsumoto_what_2021}). Elucidating the detailed effect of thermal pressure on the evolution of the expanding thin shell is our future work.

Next, let us discuss the formation of the accretion disk, which is the wind source. The accretion disk is one of the main emission sources of TDEs. Therefore, it is important to know how long it takes for the disk to form after tidal disruption for the purpose of comparing to observations. If the OUV emissions are attributed to the energy dissipation due to the debris stream-stream collision where the collision is assumed to be radiatively efficient, then the circularization time to the subsequent disk formation is estimated as 
\begin{eqnarray}
t_{\rm circ}
&&
\sim 
39\,{\rm day}\,
\left(\frac{\beta}{1.1}\right)^{-3}
\left(
\frac{
M_{\rm bh}
}{10^{6.7}\,M_\odot}
\right)^{-7/6}
\left(\frac{m_*}{M_\odot}\right)^{-1}
\left(\frac{r_*}{R_\odot}\right)^{3/2}
\nonumber
\label{eq:tcirc}
\end{eqnarray}
based on the ballistic approximation \citep{bonnerot_long-term_2017}. This is consistent with the onset time of OUV emissions, which is given by equation~(\ref{eq:td}), although how and when the accretion disk forms still remain debated. Our model with radio observations for AT2019dsg can give the constraint on the epoch of the accretion disk formation. A pivotal caveat for the argument is that if the shock region by initial debris stream-stream collision is radiatively inefficient, a collision-induced outflow (CIO) can occur \citep{lu_self-intersection_2020} instead of OUV emissions there. In this case, the disk rapidly forms in a more complicated way \citep{bonnerot_simulating_2020}, implying that the OUV peak could manifest subsequent to the disk formation. Nonetheless, our model remains applicable to this scenario given appropriate parameter modifications. Moreover, the CIO itself can be another wind source to produce the radio-emitting shell. \cite{lu_self-intersection_2020} actually discussed the radio spectra and light curves from non-Jetted, radio-emitting TDEs based on their CIO model. Our formulation (see Section 2) can be applied to exploring the evolution of the CIO-driven radio-emitting shell, although we need to consider the geometrical and energetic differences. Identifying which causes the observed radio emissions, CIO or disk wind, is a future challenge.

Finally, let us describe how it compares to other radio-emitting TDEs. Looking at the radial dependence of the density of surrounding matter in the following paper, some objects (namely ASASSN-14li \citep{alexander_discovery_2016}, CNSS J0019+00 \citep{anderson_caltech-nrao_2020}, AT2020opy \citep{goodwin_radio_2023}) follows the power-law decay similarly like AT2019dsg, while AT2019azh \citep{goodwin_at2019azh_2022} shows different unusual behavior (except for Sw J1644+57 which has an obvious relativistic jet). Our model is, therefore, applicable to the former objects, but we need sufficient observed data for comparison purposes. In our future work, we will individually examine whether our model can be applied to non-relativistic radio-emitting TDEs.
%

%
%%%%%%%%%%%%%%%%%%%
\section{Conclusions}
\label{sec:cons}
%%%%%%%%%%%%%%%%%%%
%
We have constructed a basic theoretical model of the disk-wind-driven non-relativistic expanding thin shell targeting for the radio-emitting TDEs without relativistic jets. There are five unknown parameters for our model, but most of them can be identified or constrained using the observed data. Our conclusions are summarized as follows:
\begin{enumerate}
\item
We derive an approximate solution in the form of the Taylor series. We compare it with the numerical solution in early times for the various values of the five parameters and then confirm both solutions are in good agreement consequently.

\item
We find that the expanding thin shell decelerates except for very early times. This tendency appears independently of the five parameters of our model sooner or later because the ambient matter acts against the shell's expansion.

\item
We find no single power-law of time solution for the shall radius that explains the evolution of the shell radius in early to middle times. This is because the ejecta's ram pressure significantly impacts the early time evolution of the shell radius, producing the variation of the local power-law index of time for the shell radius.

\item
We derive an asymptotic solution for the shell radius evolution at late times $R(t){\propto}t^{1/(4-s)}$, where $s$ is the power-law index of radius for the ambient matter's radial density distribution. We confirm that the numerical solutions approach the asymptotic solution at late times.

\item
Our model well explains the observed data of the shell radius evolution of AT2019dsg.
\end{enumerate}

%
%%%%%%
%\newpage
%\pagebreak
\appendix
%%%%%%
%
%\section{Appendix information}
%
%%%%%%%%%%%%%%%%%%%%%%%%%%%%%%%%%%%%%
\section{The approximate solution by the Taylor series}
\label{sec:app}
%%%%%%%%%%%%%%%%%%%%%%%%%%%%%%%%%%%%%%
%
If $y=y(x)$ is infinitely differentiable at $x_0$,  
the Taylor series yields the approximate solution of $y(x)$ around $x_0$ as
\begin{eqnarray}
y(x)
&&
={y}(x_0)
+\dot{y}(x_0)(x-x_0)
+\frac{1}{2}\ddot{y}(x_0)(x-x_0)^2
+\frac{1}{6}\dddot{y}(x_0)(x-x_0)^3
+
\mathcal{O}\big((x-x_0)^4\big)
\nonumber \\
&&
=
1 + \frac{1}{\epsilon}(x-x_0) - \frac{1}{2}\frac{\eta}{\delta\epsilon^2}(x-x_0)^2  + \frac{1}{6}\dddot{y}(x_0)(x-x_0)^3
+
\mathcal{O}\big((x-x_0)^4\big)
\label{eq:tayloreq}
\end{eqnarray}
where we find $y(x_0)=1$ and $\dot{y}(x_0)=v_0/V_0=1/\epsilon$ at $x_0=1$ and we obtain, by using equation~(\ref{eq:eom2}),
\begin{eqnarray}
\ddot{y}(x_0)
=-\frac{\eta}{\delta\epsilon^2}.
\nonumber
\label{eq:ddotx0}
\end{eqnarray}
Differentiating equation~(\ref{eq:eom2}) with respect to $x$ and subsequently evaluating it at $x=x_0=1$ 
yields
%Evaluating equation~(\ref{eq:3dots}) at $x=x_0=1$, we obtain 
\begin{eqnarray}
\dddot{y}(x_0)
=
\frac{\eta}{\delta\epsilon^3}
%\frac{\eta}{\delta}
\left(
3\frac{\eta}{\delta}
+
s-2
\right).
\label{eq:ddddotx0}
\end{eqnarray}
Substituting equation~(\ref{eq:ddddotx0}) into equation~(\ref{eq:tayloreq}), we obtain the approximate equation of $y(x)$ near $x=x_0$ by the third order of $x$ as
\begin{equation}
y(x)
=
%\approx
1 + \frac{1}{\epsilon}(x-x_0) - \frac{1}{2}\frac{\eta}{\delta\epsilon^2}(x-x_0)^2  + \frac{1}{6}\frac{\eta}{\delta\epsilon^3}
\left[
3\frac{\eta}{\delta}
+
s-2
\right]
(x-x_0)^3
+\cdots
\end{equation}
%

%
%%%%%%%%%%%%%%
\section{Acknowledgments}.
\begin{acknowledgments}
%We acknowledge anonymous referee for helpful comments.
We thank Shuta J. Tanaka for his useful comments. We also thank the anonymous referee for helpful comments, which improved the content and clarity of the paper. K.H. acknowledges the Institute for Theory and Computation, Harvard-Smithsonian Center for Astrophysics, for the warm hospitality during the sabbatical year. K.H. has been supported by the Basic Science Research Program through the National Research Foundation of Korea (NRF) funded by the Ministry of Education (2016R1A5A1013277 and 2020R1A2C1007219). This work was financially supported by the Research Year of Chungbuk National University in 2021 and was also supported in part by the National Science Foundation under Grant No. NSF PHY-1748958. R.Y. has been supported by JSPS KAKENHI (Grant Nos. 20H01881, 22H00119, 22H01251, 23H01211, and 23H04899).
\end{acknowledgments}
%%%%%%%%%%%%%%
%

%
%%%%%%%%%%%%%%%
\bibliography{khry2023}{}

\begin{thebibliography}{}
\expandafter\ifx\csname natexlab\endcsname\relax\def\natexlab#1{#1}\fi
\providecommand{\url}[1]{\href{#1}{#1}}
\providecommand{\dodoi}[1]{doi:~\href{http://doi.org/#1}{\nolinkurl{#1}}}
\providecommand{\doeprint}[1]{\href{http://ascl.net/#1}{\nolinkurl{http://ascl.net/#1}}}
\providecommand{\doarXiv}[1]{\href{https://arxiv.org/abs/#1}{\nolinkurl{https://arxiv.org/abs/#1}}}

\bibitem[{Alexander {et~al.}(2016)Alexander, Berger, Guillochon, Zauderer, \&
  Williams}]{alexander_discovery_2016}
Alexander, K.~D., Berger, E., Guillochon, J., Zauderer, B.~A., \& Williams, P.
  K.~G. 2016, The Astrophysical Journal, 819, L25,
  \dodoi{10.3847/2041-8205/819/2/L25}

\bibitem[{Alexander {et~al.}(2020)Alexander, van Velzen, Horesh, \&
  Zauderer}]{alexander_radio_2020}
Alexander, K.~D., van Velzen, S., Horesh, A., \& Zauderer, B.~A. 2020, Space
  Sci Rev, 216, 81, \dodoi{10.1007/s11214-020-00702-w}

\bibitem[{Alexander {et~al.}(2017)Alexander, Wieringa, Berger, Saxton, \&
  Komossa}]{alexander_radio_2017}
Alexander, K.~D., Wieringa, M.~H., Berger, E., Saxton, R.~D., \& Komossa, S.
  2017, The Astrophysical Journal, 837, 153, \dodoi{10.3847/1538-4357/aa6192}

\bibitem[{Anderson {et~al.}(2020)Anderson, Mooley, Hallinan, Dong, Phinney,
  Horesh, Bourke, Cenko, Frail, Kulkarni, \&
  Myers}]{anderson_caltech-nrao_2020}
Anderson, M.~M., Mooley, K.~P., Hallinan, G., {et~al.} 2020, The Astrophysical
  Journal, 903, 116, \dodoi{10.3847/1538-4357/abb94b}

\bibitem[{Andreoni {et~al.}(2022)Andreoni, Coughlin, Perley, Yao, Lu, Cenko,
  Kumar, Anand, Ho, Kasliwal, de~Ugarte~Postigo, Sagu^^c3^^a9s-Carracedo,
  Schulze, Kann, Kulkarni, Sollerman, Tanvir, Rest, Izzo, Somalwar, Kaplan,
  Ahumada, Anupama, Auchettl, Barway, Bellm, Bhalerao, Bloom, Bremer, Bulla,
  Burns, Campana, Chandra, Charalampopoulos, Cooke, D'Elia, Das, Dobie,
  Ag^^c3^^bc^^c3^^ad~Fern^^c3^^a1ndez, Freeburn, Fremling, Gezari, Goode,
  Graham, Hammerstein, Karambelkar, Kilpatrick, Kool, Krips, Laher, Leloudas,
  Levan, Lundquist, Mahabal, Medford, Miller, M^^c3^^b6ller, Mooley, Nayana,
  Nir, Pang, Paraskeva, Perley, Petitpas, Pursiainen, Ravi, Ridden-Harper,
  Riddle, Rigault, Rodriguez, Rusholme, Sharma, Smith, Stein, Th^^c3^^b6ne,
  Tohuvavohu, Valdes, van Roestel, Vergani, Wang, \&
  Zhang}]{andreoni_very_2022}
Andreoni, I., Coughlin, M.~W., Perley, D.~A., {et~al.} 2022, Nature, 612, 430,
  \dodoi{10.1038/s41586-022-05465-8}

\bibitem[{Barniol~Duran {et~al.}(2013)Barniol~Duran, Nakar, \&
  Piran}]{barniol_duran_radius_2013}
Barniol~Duran, R., Nakar, E., \& Piran, T. 2013, The Astrophysical Journal,
  772, 78, \dodoi{10.1088/0004-637X/772/1/78}

\bibitem[{Bloom {et~al.}(2011)Bloom, Giannios, Metzger, Cenko, Perley, Butler,
  Tanvir, Levan, O'Brien, Strubbe, De~Colle, Ramirez-Ruiz, Lee, Nayakshin,
  Quataert, King, Cucchiara, Guillochon, Bower, Fruchter, Morgan, \& van~der
  Horst}]{bloom_possible_2011}
Bloom, J.~S., Giannios, D., Metzger, B.~D., {et~al.} 2011, Science, 333, 203,
  \dodoi{10.1126/science.1207150}

\bibitem[{Bonnerot \& Lu(2020)}]{bonnerot_simulating_2020}
Bonnerot, C., \& Lu, W. 2020, Monthly Notices of the Royal Astronomical
  Society, 495, 1374, \dodoi{10.1093/mnras/staa1246}

\bibitem[{Bonnerot {et~al.}(2017)Bonnerot, Rossi, \&
  Lodato}]{bonnerot_long-term_2017}
Bonnerot, C., Rossi, E.~M., \& Lodato, G. 2017, Monthly Notices of the Royal
  Astronomical Society, 464, 2816, \dodoi{10.1093/mnras/stw2547}

\bibitem[{Bonnerot {et~al.}(2016)Bonnerot, Rossi, Lodato, \&
  Price}]{bonnerot_disc_2016}
Bonnerot, C., Rossi, E.~M., Lodato, G., \& Price, D.~J. 2016, Monthly Notices
  of the Royal Astronomical Society, 455, 2253, \dodoi{10.1093/mnras/stv2411}

\bibitem[{Brown {et~al.}(2015)Brown, Levan, Stanway, Tanvir, Cenko, Berger,
  Chornock, \& Cucchiaria}]{brown_swift_2015}
Brown, G.~C., Levan, A.~J., Stanway, E.~R., {et~al.} 2015, Monthly Notices of
  the Royal Astronomical Society, 452, 4297, \dodoi{10.1093/mnras/stv1520}

\bibitem[{Burrows {et~al.}(2011)Burrows, Kennea, Ghisellini, Mangano, Zhang,
  Page, Eracleous, Romano, Sakamoto, Falcone, Osborne, Campana, Beardmore,
  Breeveld, Chester, Corbet, Covino, Cummings, D'Avanzo, D'Elia, Esposito,
  Evans, Fugazza, Gelbord, Hiroi, Holland, Huang, Im, Israel, Jeon, Jeon, Jun,
  Kawai, Kim, Krimm, Marshall, {P. M^^c3^^a9sz^^c3^^a1ros}, Negoro, Omodei,
  Park, Perkins, Sugizaki, Sung, Tagliaferri, Troja, Ueda, Urata, Usui,
  Antonelli, Barthelmy, Cusumano, Giommi, Melandri, Perri, Racusin, Sbarufatti,
  Siegel, \& Gehrels}]{burrows_relativistic_2011}
Burrows, D.~N., Kennea, J.~A., Ghisellini, G., {et~al.} 2011, Nature, 476, 421,
  \dodoi{10.1038/nature10374}

\bibitem[{Cannizzo {et~al.}(1990)Cannizzo, Lee, \&
  Goodman}]{cannizzo_disk_1990}
Cannizzo, J.~K., Lee, H.~M., \& Goodman, J. 1990, The Astrophysical Journal,
  351, 38, \dodoi{10.1086/168442}

\bibitem[{Cendes {et~al.}(2021)Cendes, Alexander, Berger, Eftekhari, Williams,
  \& Chornock}]{cendes_radio_2021}
Cendes, Y., Alexander, K.~D., Berger, E., {et~al.} 2021, The Astrophysical
  Journal, 919, 127, \dodoi{10.3847/1538-4357/ac110a}

\bibitem[{Cendes {et~al.}(2022)Cendes, Berger, Alexander, Gomez, Hajela,
  Chornock, Laskar, Margutti, Metzger, Bietenholz, Brethauer, \&
  Wieringa}]{cendes_mildly_2022}
Cendes, Y., Berger, E., Alexander, K., {et~al.} 2022, A {Mildly} {Relativistic}
  {Outflow} {Launched} {Two} {Years} after {Disruption} in the {Tidal}
  {Disruption} {Event} {AT2018hyz}, Tech. rep.
\newblock \url{https://ui.adsabs.harvard.edu/abs/2022arXiv220614297C}

\bibitem[{Cenko {et~al.}(2012)Cenko, Krimm, Horesh, Rau, Frail, Kennea, Levan,
  Holland, Butler, Quimby, Bloom, Filippenko, Gal-Yam, Greiner, Kulkarni, Ofek,
  Olivares~E., Schady, Silverman, Tanvir, \& Xu}]{cenko_swift_2012}
Cenko, S.~B., Krimm, H.~A., Horesh, A., {et~al.} 2012, The Astrophysical
  Journal, 753, 77, \dodoi{10.1088/0004-637X/753/1/77}

\bibitem[{Chevalier(1982)}]{chevalier_radio_1982}
Chevalier, R.~A. 1982, The Astrophysical Journal, 259, 302,
  \dodoi{10.1086/160167}

\bibitem[{Chevalier(1998)}]{chevalier_synchrotron_1998}
---. 1998, The Astrophysical Journal, 499, 810, \dodoi{10.1086/305676}

\bibitem[{Dai {et~al.}(2018)Dai, McKinney, Roth, Ramirez-Ruiz, \&
  Miller}]{dai_unified_2018}
Dai, L., McKinney, J.~C., Roth, N., Ramirez-Ruiz, E., \& Miller, M.~C. 2018,
  The Astrophysical Journal, 859, L20, \dodoi{10.3847/2041-8213/aab429}

\bibitem[{Evans \& Kochanek(1989)}]{evans_tidal_1989}
Evans, C.~R., \& Kochanek, C.~S. 1989, The Astrophysical Journal, 346, L13,
  \dodoi{10.1086/185567}

\bibitem[{Goodwin {et~al.}(2022)Goodwin, van Velzen, Miller-Jones, Mummery,
  Bietenholz, Wederfoort, Hammerstein, Bonnerot, Hoffmann, \&
  Yan}]{goodwin_at2019azh_2022}
Goodwin, A.~J., van Velzen, S., Miller-Jones, J. C.~A., {et~al.} 2022, Monthly
  Notices of the Royal Astronomical Society, 511, 5328,
  \dodoi{10.1093/mnras/stac333}

\bibitem[{Goodwin {et~al.}(2023)Goodwin, Miller-Jones, van Velzen, Bietenholz,
  Greenland, Cenko, Gezari, Horesh, Sivakoff, Yan, Yu, \&
  Zhang}]{goodwin_radio_2023}
Goodwin, A.~J., Miller-Jones, J. C.~A., van Velzen, S., {et~al.} 2023, Monthly
  Notices of the Royal Astronomical Society, 518, 847,
  \dodoi{10.1093/mnras/stac3127}

\bibitem[{Guillochon {et~al.}(2016)Guillochon, McCourt, Chen, Johnson, \&
  Berger}]{guillochon_unbound_2016}
Guillochon, J., McCourt, M., Chen, X., Johnson, M.~D., \& Berger, E. 2016, The
  Astrophysical Journal, 822, 48, \dodoi{10.3847/0004-637X/822/1/48}

\bibitem[{Hayasaki \& Jonker(2021)}]{hayasaki_origin_2021}
Hayasaki, K., \& Jonker, P.~G. 2021, The Astrophysical Journal, 921, 20,
  \dodoi{10.3847/1538-4357/ac18c2}

\bibitem[{Hayasaki {et~al.}(2013)Hayasaki, Stone, \&
  Loeb}]{hayasaki_finite_2013}
Hayasaki, K., Stone, N., \& Loeb, A. 2013, Monthly Notices of the Royal
  Astronomical Society, 434, 909, \dodoi{10.1093/mnras/stt871}

\bibitem[{Hayasaki {et~al.}(2016)Hayasaki, Stone, \&
  Loeb}]{hayasaki_circularization_2016}
---. 2016, Monthly Notices of the Royal Astronomical Society, 461, 3760,
  \dodoi{10.1093/mnras/stw1387}

\bibitem[{Hills(1975)}]{hills_possible_1975}
Hills, J.~G. 1975, Nature, 254, 295, \dodoi{10.1038/254295a0}

\bibitem[{Jiang {et~al.}(2016)Jiang, Guillochon, \& Loeb}]{jiang_prompt_2016}
Jiang, Y.-F., Guillochon, J., \& Loeb, A. 2016, The Astrophysical Journal, 830,
  125, \dodoi{10.3847/0004-637X/830/2/125}

\bibitem[{Krolik {et~al.}(2016)Krolik, Piran, Svirski, \&
  Cheng}]{krolik_asassn-14li_2016}
Krolik, J., Piran, T., Svirski, G., \& Cheng, R.~M. 2016, The Astrophysical
  Journal, 827, 127, \dodoi{10.3847/0004-637X/827/2/127}

\bibitem[{Lei {et~al.}(2016)Lei, Yuan, Zhang, \& Wang}]{lei_igr_2016}
Lei, W.-H., Yuan, Q., Zhang, B., \& Wang, D. 2016, The Astrophysical Journal,
  816, 20, \dodoi{10.3847/0004-637X/816/1/20}

\bibitem[{Levan {et~al.}(2011)Levan, Tanvir, Cenko, Perley, Wiersema, Bloom,
  Fruchter, de~Ugarte~Postigo, O'Brien, Butler, van~der Horst, Leloudas,
  Morgan, Misra, Bower, Farihi, Tunnicliffe, Modjaz, Silverman, Hjorth,
  Th^^c3^^b6ne, Cucchiara, Cer^^c3^^b3n, Castro-Tirado, Arnold, Bremer, Brodie,
  Carroll, Cooper, Curran, Cutri, Ehle, Forbes, Fynbo, Gorosabel, Graham,
  Hoffman, Guziy, Jakobsson, Kamble, Kerr, Kasliwal, Kouveliotou, Kocevski,
  Law, Nugent, Ofek, Poznanski, Quimby, Rol, Romanowsky,
  S^^c3^^a1nchez-Ram^^c3^^adrez, Schulze, Singh, van Spaandonk, Starling,
  Strom, Tello, Vaduvescu, Wheatley, Wijers, Winters, \&
  Xu}]{levan_extremely_2011}
Levan, A.~J., Tanvir, N.~R., Cenko, S.~B., {et~al.} 2011, Science, 333, 199,
  \dodoi{10.1126/science.1207143}

\bibitem[{Longair(2011)}]{longair_high_2011}
Longair, M.~S. 2011, High {Energy} {Astrophysics}.
\newblock \url{https://ui.adsabs.harvard.edu/abs/2011hea..book.....L}

\bibitem[{Lu \& Bonnerot(2020)}]{lu_self-intersection_2020}
Lu, W., \& Bonnerot, C. 2020, Monthly Notices of the Royal Astronomical
  Society, 492, 686, \dodoi{10.1093/mnras/stz3405}

\bibitem[{Matsumoto {et~al.}(2021)Matsumoto, Piran, \&
  Krolik}]{matsumoto_what_2021}
Matsumoto, T., Piran, T., \& Krolik, J.~H. 2021, arXiv:2109.02648 [astro-ph].
\newblock \url{http://arxiv.org/abs/2109.02648}

\bibitem[{Mattila {et~al.}(2018)Mattila, P^^c3^^a9rez-Torres, Efstathiou,
  Mimica, Fraser, Kankare, Alberdi, Aloy, Heikkil^^c3^^a4, Jonker, Lundqvist,
  Mart^^c3^^ad-Vidal, Meikle, Romero-Ca^^c3^^b1izales, Smartt, Tsygankov,
  Varenius, Alonso-Herrero, Bondi, Fransson, Herrero-Illana, Kangas, Kotak,
  Ram^^c3^^adrez-Olivencia, V^^c3^^a4is^^c3^^a4nen, Beswick, Clements, Greimel,
  Harmanen, Kotilainen, Nandra, Reynolds, Ryder, Walton, Wiik, \&
  ^^c3^^96stlin}]{mattila_dust-enshrouded_2018}
Mattila, S., P^^c3^^a9rez-Torres, M., Efstathiou, A., {et~al.} 2018, Science,
  361, 482, \dodoi{10.1126/science.aao4669}

\bibitem[{Mummery \& Balbus(2020)}]{mummery_spectral_2020}
Mummery, A., \& Balbus, S.~A. 2020, Monthly Notices of the Royal Astronomical
  Society, 492, 5655, \dodoi{10.1093/mnras/staa192}

\bibitem[{Pasham \& van Velzen(2018)}]{pasham_discovery_2018}
Pasham, D.~R., \& van Velzen, S. 2018, The Astrophysical Journal, 856, 1,
  \dodoi{10.3847/1538-4357/aab361}

\bibitem[{Rees(1988)}]{rees_tidal_1988}
Rees, M.~J. 1988, Nature, 333, 523, \dodoi{10.1038/333523a0}

\bibitem[{Shiokawa {et~al.}(2015)Shiokawa, Krolik, Cheng, Piran, \&
  Noble}]{shiokawa_general_2015}
Shiokawa, H., Krolik, J.~H., Cheng, R.~M., Piran, T., \& Noble, S.~C. 2015, The
  Astrophysical Journal, 804, 85, \dodoi{10.1088/0004-637X/804/2/85}

\bibitem[{Somalwar {et~al.}(2023)Somalwar, Ravi, Dong, Chen, Breen, Chandra,
  Clarke, De, Gaensler, Hallinan, Laha, Law, Myers, Parsotan, Peters, \&
  Polisensky}]{somalwar_candidate_2023}
Somalwar, J.~J., Ravi, V., Dong, D.~Z., {et~al.} 2023, The Astrophysical
  Journal, 945, 142, \dodoi{10.3847/1538-4357/acbafc}

\bibitem[{Stein {et~al.}(2021)Stein, Velzen, Kowalski, Franckowiak, Gezari,
  Miller-Jones, Frederick, Sfaradi, Bietenholz, Horesh, Fender, Garrappa,
  Ahumada, Andreoni, Belicki, Bellm, B^^c3^^b6ttcher, Brinnel, Burruss, Cenko,
  Coughlin, Cunningham, Drake, Farrar, Feeney, Foley, Gal-Yam, Golkhou, Goobar,
  Graham, Hammerstein, Helou, Hung, Kasliwal, Kilpatrick, Kong, Kupfer, Laher,
  Mahabal, Masci, Necker, Nordin, Perley, Rigault, Reusch, Rodriguez,
  Rojas-Bravo, Rusholme, Shupe, Singer, Sollerman, Soumagnac, Stern, Taggart,
  van Santen, Ward, Woudt, \& Yao}]{stein_tidal_2021}
Stein, R., Velzen, S.~v., Kowalski, M., {et~al.} 2021, Nat Astron, 5, 510,
  \dodoi{10.1038/s41550-020-01295-8}

\bibitem[{Stone {et~al.}(2020)Stone, Vasiliev, Kesden, Rossi, Perets, \&
  Amaro-Seoane}]{stone_rates_2020}
Stone, N.~C., Vasiliev, E., Kesden, M., {et~al.} 2020, Space Sci Rev, 216, 35,
  \dodoi{10.1007/s11214-020-00651-4}

\bibitem[{Strubbe \& Quataert(2009)}]{strubbe_optical_2009}
Strubbe, L.~E., \& Quataert, E. 2009, Monthly Notices of the Royal Astronomical
  Society, 400, 2070, \dodoi{10.1111/j.1365-2966.2009.15599.x}

\bibitem[{van Velzen {et~al.}(2016)van Velzen, Anderson, Stone, Fraser, Wevers,
  Metzger, Jonker, van~der Horst, Staley, Mendez, Miller-Jones, Hodgkin,
  Campbell, \& Fender}]{van_velzen_radio_2016}
van Velzen, S., Anderson, G.~E., Stone, N.~C., {et~al.} 2016, Science, 351, 62,
  \dodoi{10.1126/science.aad1182}

\bibitem[{Yalinewich {et~al.}(2019)Yalinewich, Steinberg, Piran, \&
  Krolik}]{yalinewich_radio_2019}
Yalinewich, A., Steinberg, E., Piran, T., \& Krolik, J.~H. 2019, Monthly
  Notices of the Royal Astronomical Society, 487, 4083,
  \dodoi{10.1093/mnras/stz1567}

\bibitem[{Zauderer {et~al.}(2011)Zauderer, Berger, Soderberg, Loeb, Narayan,
  Frail, Petitpas, Brunthaler, Chornock, Carpenter, Pooley, Mooley, Kulkarni,
  Margutti, Fox, Nakar, Patel, Volgenau, Culverhouse, Bietenholz, Rupen,
  Max-Moerbeck, Readhead, Richards, Shepherd, Storm, \&
  Hull}]{zauderer_birth_2011}
Zauderer, B.~A., Berger, E., Soderberg, A.~M., {et~al.} 2011, Nature, 476, 425,
  \dodoi{10.1038/nature10366}

\end{thebibliography}
\bibliographystyle{aasjournal}
%%%%%%%%%%%%%%%
%

\end{document}